\documentclass[12pt,english]{article}

\usepackage{helvet}

\usepackage[T1]{fontenc}
\usepackage[latin9]{inputenc}
\usepackage{geometry}
\usepackage{amsbsy}
\usepackage{amstext}
\usepackage{graphicx}
\usepackage{setspace}
\makeatletter
\date{}
\usepackage[font={sf, stretch=1.3}]{caption}
\usepackage{xurl}

\makeatother

\usepackage{babel}
\begin{document}
\title{Cognition of time and thinkings beyond}
\author{$\text{Zedong Bi}^{1,2,3*}$ }

\maketitle
$^{1}$Lingang Laboratory, Shanghai 200031, China

$^{2}$Institute for Future, Qingdao University, Qingdao 266071, China 

$^{3}$School of Automation, Shandong Key Laboratory of Industrial
Control Technology, Qingdao University, Qingdao 266071, China 

{*}zedong.bi@outlook.com 

\pagebreak 

\section*{Abstract}

A pervasive research protocol of cognitive neuroscience is to train
subjects to perform deliberately designed experiments and record brain
activity simultaneously, aiming to understand the brain mechanism
underlying cognition. However, how the results of this protocol of
research can be applied in technology is seldomly discussed. Here,
I review the studies on time processing of the brain as examples of
this research protocol, as well as two main application areas of neuroscience
(neuroengineering and brain-inspired artificial intelligence). Time
processing is an indispensable dimension of cognition, and time is
also an indispensable dimension of any real-world signal to be processed
in technology. Therefore, one may expect that the studies of time
processing in cognition profoundly influence brain-related technology.
Surprisingly, I found that the results from cognitive studies on timing
processing are hardly helpful in solving practical problems. This
awkward situation may be due to the lack of generalizability of the
results of cognitive studies, which are under well-controlled laboratory
conditions, to real-life situations. This lack of generalizability
may be rooted in the fundamental unknowability of the world (including
cognition). Overall, this paper questions and criticizes the usefulness
and prospect of the above-mentioned research protocol of cognitive
neuroscience. I then give three suggestions for future research. First,
to improve the generalizability of research, it is better to study
brain activity under real-life conditions instead of in well-controlled
laboratory experiments. Second, to overcome the unknowability of the
world, we can engineer an easily accessible surrogate of the object
under investigation, so that we can predict the behavior of the object
under investigation by experimenting on the surrogate. Third, I call
for technology-oriented research, with the aim of technology creation
instead of knowledge discovery. 

\pagebreak 

\section*{Introduction}

Atomism, the idea that the universe is composed of fundamental components
known as atoms, is perhaps the most influential philosophy leading
scientific research. Richard Feynman regarded atomism as the most
important thinking we should pass to the next generation \cite{Feynman_2011}:
various physical changes and chemical reactions can be readily explained
by supposing the movements and interactions of atoms \cite{Feynman_2011}. 

\begin{figure}
\center \includegraphics[scale=0.8]{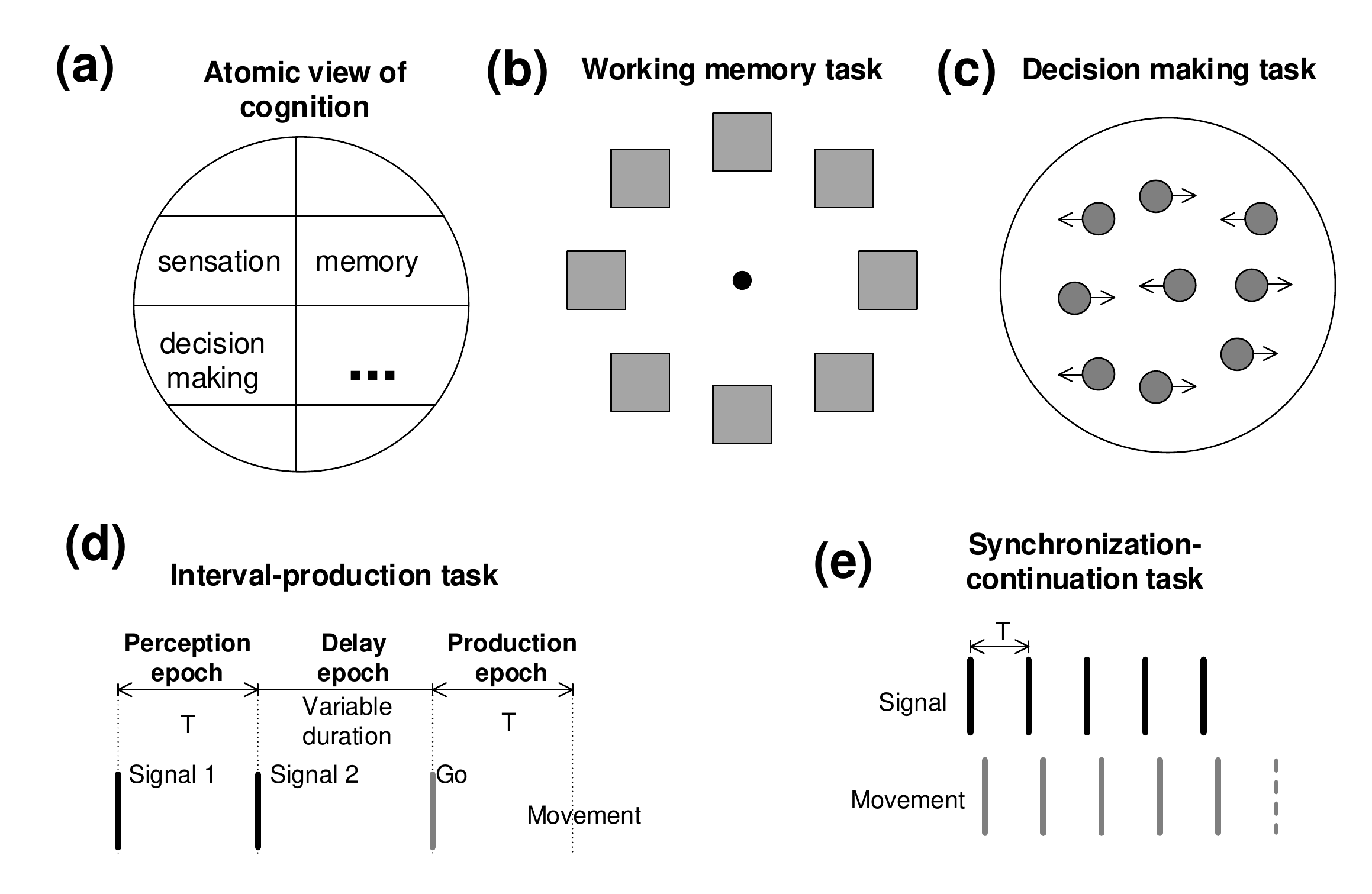}

\caption{\textbf{\label{fig:basic-cog-studies}Illustration of basic cognitive
studies.} (\textbf{a}) Basic cognitive studies are guided by the philosophy
of atomism, so that cognition is divided into many elements, with
each element separately studied. Atomists believe that we can understand
the whole of cognition after understanding each element. (\textbf{b})
A classical experiment to study working memory. The subject fixated
on a central point, and a visuospatial cue (one of the eight grey
boxes) was presented briefly, followed by a mnemonic delay. After
the delay, the subject was to make a saccadic eye movement to the
remembered location. (\textbf{c}) A classical experiment to study
decision making. There were two types of random dots, one type moving
leftward, the other moving rightward. The subject was to make the
decision on which type had more dots. (\textbf{d}) Schematic of the
time production task. The subject received two signals (black bars)
separated by time interval $T$; after a delay epoch with variable
duration, a go cue (grey bar) appeared, and the subject was to move
at time $T$ after the go cue. (\textbf{e}) Schematic of the synchronization-continuation
task. The subject was to move (grey bars) immediately following a
sequence of signals (black bars) with period $T$. The subject was
still to move with period $T$ (dashed bar) after the signal was removed. }

\end{figure}

Atomism also strongly influences cognitive neuroscience. Psychologists
have divided cognition into several elements: perception, learning,
memory, decision making, etc \cite{Baldwin_1893}. Single elements
can be further divided into several sub-elements from different angles.
For example, perception can be divided into the perception of space
and the perception of time, or into the perception of visual information
and auditory information; memory can be divided into short-term and
long-term memory, or into episodic and semantic memory, etc. By investigating
the brain activity when the subject is performing each element of
cognition, neuroscientists aim to understand the biological backend
of cognition by collecting all these pieces together (\textbf{Figure
\ref{fig:basic-cog-studies}a}). From this atomistic perspective,
studying a single cognitive element is the foundation for understanding
cognition, so I named this research protocol to be \textit{basic}. 

To perform basic cognitive studies, researchers elaborately design
simple and well-controlled experimental conditions to study a single
cognitive element while teasing apart the influence from other elements.
For example, to study working memory, researchers trained monkeys
to recall a visual cue after a delay period \cite{Constantinidis_2001}
(\textbf{Figure \ref{fig:basic-cog-studies}b}). To study decision
making, researchers trained monkeys to watch two types of dots moving
toward oppositive directions and then decide which type had more dots
\cite{Roitman_2002} (\textbf{Figure \ref{fig:basic-cog-studies}c}).
The study of time cognition also stems from atomism. To focus on the
processing of time while disentangling other cognitive elements (such
as the perception of spatial information), psychologists or neuroscientists
train subjects to perform simple but deliberately designed timing
tasks. In a classical experiment \cite{Rakitin_1998}, participants
were presented with specific time intervals delimited by stimuli,
and then were asked to reproduce the interval (\textbf{Figure \ref{fig:basic-cog-studies}d}).
When subjects were performing these simple and deliberately-designed
tasks, their brain activities were recorded. In this way, researchers
could propose neural-network mechanisms underpinning basic elements
of cognition. 

Perhaps different scientists have different beliefs about the ultimate
task of science. Some people believe that the pure aim of science
is to satisfy our curiosity about the world. But in my opinion, scientific
results must be implemented in technology and benefit the mass of
people before scientific results complete their mission. However,
the status and prospects of the technological applications of basic
cognitive studies have seldomly been discussed. In this paper, I will
discuss technological applications of basic cognitive studies. I will
start this discussion by reviewing cognitive studies of time processing
of the brain (i.e., basic timing studies) as examples of basic cognitive
studies. Time processing is an indispensable dimension of cognition
\cite{Merchant_2013}, and time is also an indispensable dimension
of any real-world signal to be processed by technology. Therefore,
one may expect that the results of basic timing studies lay down the
foundations for processing temporal signals in brain-related technology.
Unfortunately, after reviewing two fields of brain-related technology,
neuroengineering for brain health and brain-inspired artificial intelligence,
which are two promising application fields of neuroscience suggested
by the China brain project \cite{Poo_2016}, I found that the results
of basic timing studies are hardly helpful in solving practical problems. 

\begin{figure}
\center \includegraphics{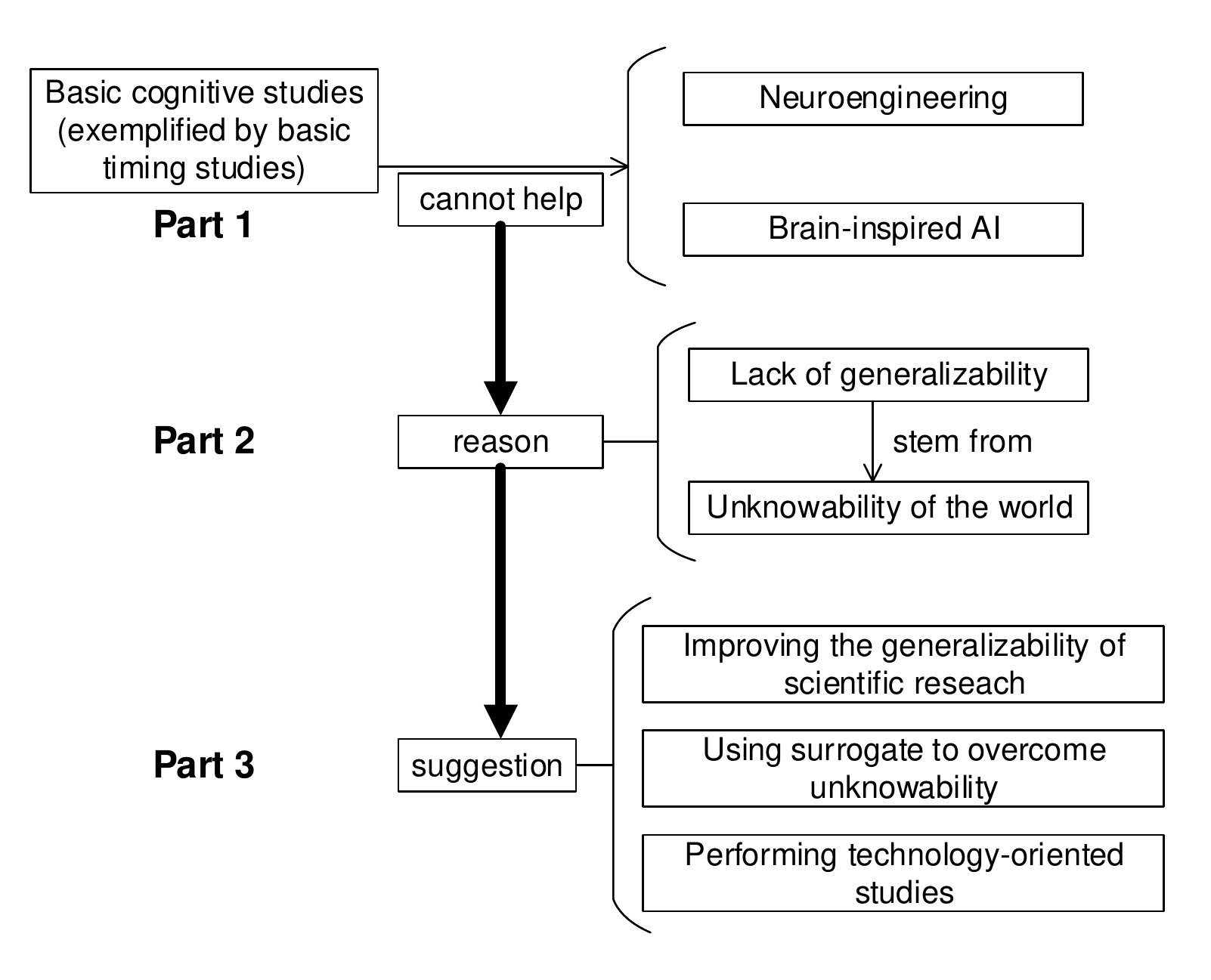}

\caption{\textbf{\label{fig:Overview}Overview of this paper. }I first (Part
1) review main results of basic timing studies, and two application
fields (neuroengineering and brain-inspired AI), showing that basic
timing studies (perhaps more generally, basic cognitive studies) cannot
help the application fields of neuroscience. Then (Part 2) I propose
that this awkward situation is due to the lack of generalizability
of the basic timing studies and, more fundamentally, the unknowability
of the world. At last (Part 3), I suggest researchers improve the
generalizability of their results, engineer surrogates to overcome
the unknowability of the world, and perform technology-oriented studies. }

\end{figure}

I then try to explain this awkward situation and give suggestions
for future research (\textbf{Figure \ref{fig:Overview}}). In my opinion,
the difficulty of basic timing studies (or more generally, basic cognitive
studies) in technological applications is closely related to their
lack of generalizability. In other words, the results of basic timing
studies depend on the specific conditions and tasks in the laboratory
experiments that found these results, but these results may not be
correct in other situations. This lack of generalizability may fundamentally
lie in the unknowability of the world (including cognition). In other
words, the capability of knowledge to describe the world is fundamentally
limited, so the generalizability of our knowledge to various situations
in the world is fundamentally limited, and therefore, the capability
of knowledge to guide technological creation to change the world is
also fundamentally limited. 

I then give three suggestions for future research (\textbf{Figure
\ref{fig:Overview}}). To improve the generalizability of results,
I suggest researchers analyze brain activities when the animals are
in real life instead of performing simple tasks in well-controlled
experimental conditions, and examine their obtained results under
various situations. To deal with the unknowability of the world, I
suggest researchers engineer surrogates of the object under investigation,
so that we can easily predict the behavior of the investigated object
using the surrogate even without understanding how the object under
investigation works. At last, due to the fundamental unknowability
of the world, I suggest researchers perform technology-oriented research
with the aim of technology creation, instead of science-oriented research
with the aim of knowledge discovery. 

\section*{Basic timing studies}

This section will give a brief overview of basic timing studies. There
are mainly two paradigms to study time cognition. The first paradigm
is interval timing, in which the subject is trained to perceive or
produce a single time interval (\textbf{Figure \ref{fig:basic-cog-studies}d}).
The second paradigm is beat timing, in which the subject is trained
to perceive or produce a sequence of time intervals rhythmically (\textbf{Figure
\ref{fig:basic-cog-studies}e}). A classical interval-timing task
is the interval-production task \cite{Rakitin_1998}, in which a subject
is presented with a specific time interval delimited by stimuli, and
then is asked to reproduce the interval (\textbf{Figure \ref{fig:basic-cog-studies}d}).
A classical beat-timing task is the synchronization-and-continuation
task \cite{Gamez_2019}, in which a subject is required to act following
a sequence of rhythmic stimuli, and continue to act rhythmically after
the removal of the stimuli (\textbf{Figure \ref{fig:basic-cog-studies}e}).
By recording brain activities when subjects are performing these tasks,
researchers can discover features of brain dynamics related to these
tasks, thereby getting insight into the neural-network mechanisms
underpinning time cognition.

\begin{figure}
\center \includegraphics[scale=0.8]{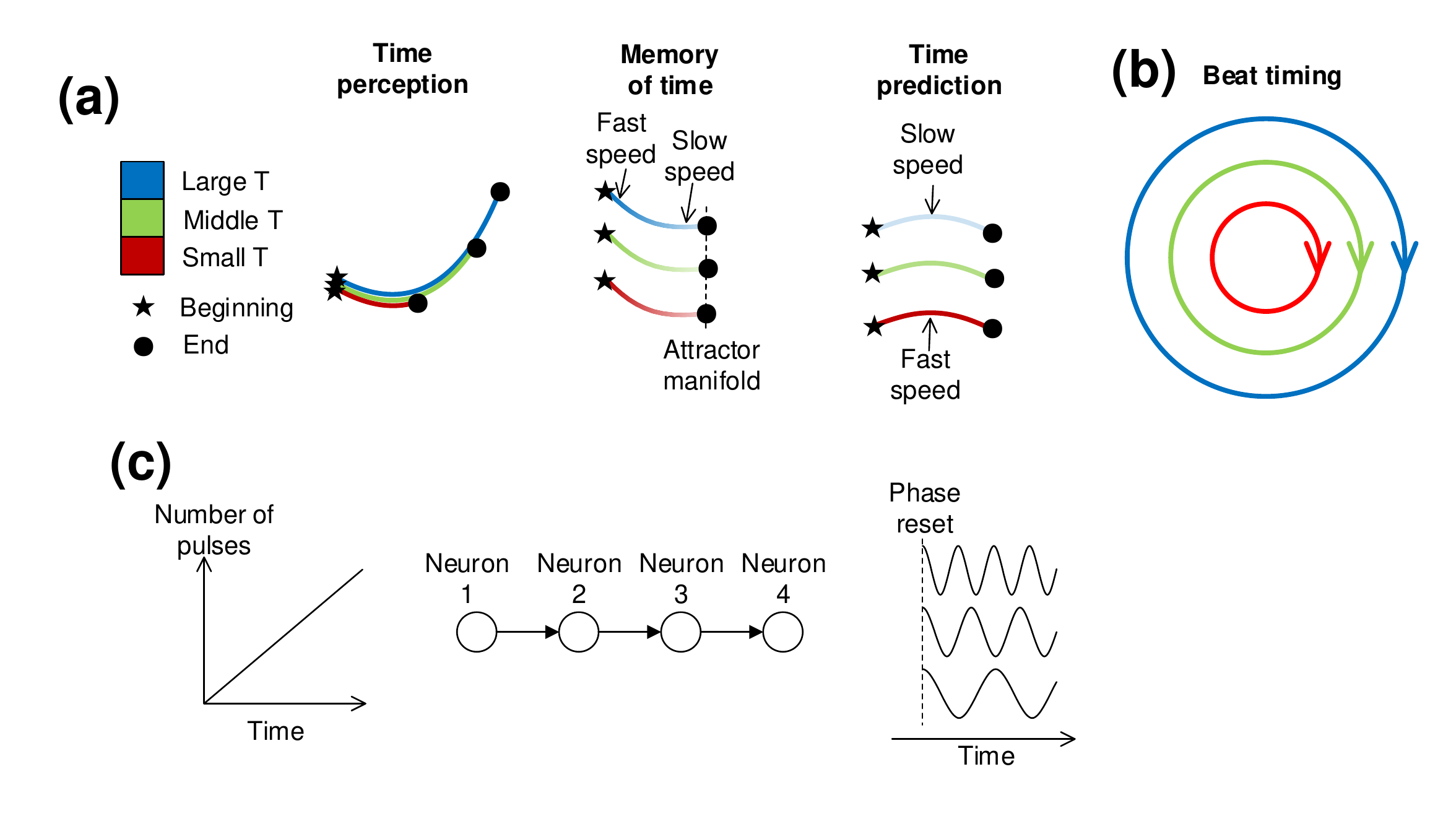}

\caption{\textbf{\label{fig:basic_timing_model} Some results of basic timing
studies.} (\textbf{a}) Dynamic features of neural networks in the
interval-production task. Left: For time perception in the perception
epoch (see \textbf{Figure \ref{fig:basic-cog-studies}d}), the network
exhibits a stereotypical trajectory whose final position determines
the perceived time interval $T$ (see \textbf{Figure \ref{fig:basic-cog-studies}d}).
Lines with blue, green and red colors respectively represent the trajectories
when $T$ is large, middle and small. Asterisk and circle respectively
represent the beginning and end of the trajectory. Middle: In the
delay epoch, time intervals are maintained in the working memory as
positions (black dots) in an attractor manifold. The speed of state
evolution with time decreases near the attractor (indicated by the
lighter color near the attractor). Right: In the production epoch,
time prediction is performed when the network state evolves along
isomorphic trajectories, with the speed of state evolution inversely
scaling with the to-be-produced interval $T$. (\textbf{b}) Dynamic
features of neural networks in the synchronization-continuation task.
With different periods $T$ (see \textbf{Figure \ref{fig:basic-cog-studies}e}),
the network state evolves along different circular trajectories at
the same speed, but the radius of the circular trajectory increases
with $T$. (\textbf{c}) Some computational models for the neural mechanisms
of time sensing. Left: In the pacemaker-accumulator model, time is
measured by the accumulated number of pulses emitted from the pacemaker.
Middle: In the synfire chain model, time can also be measured by the
sequential firing of a chain of neurons. Right: In the beat-frequency
model, time is measured by the activity pattern of a group of oscillators
with heterogeneous frequencies after phase resetting. }

\end{figure}

In the interval-production task (\textbf{Figure \ref{fig:basic-cog-studies}d}),
a neural network perceives time by evolving its state along a stereotypical
trajectory in the perception epoch, maintains time intervals in working
memory using a manifold of line attractor in the delay epoch, and
predicts a coming event by evolving its state along isomorphic trajectories
with the speed of state evolution inversely scaling with the to-be-produced
time interval in the production epoch \cite{Bi_2020} (\textbf{Figure
\ref{fig:basic_timing_model}a}). These dynamic features are consistent
with the experimental findings in other interval-timing tasks \cite{Jin_2009,Mita_2009,Wang_2018}.
In the synchronization-continuation task (\textbf{Figure \ref{fig:basic-cog-studies}e}),
different beating periods $T$ are encoded by circular trajectories
\cite{Gamez_2019} (\textbf{Figure \ref{fig:basic_timing_model}b}).
The radii of these circular trajectories increase with the period
$T$, but the speed of evolution of the network state with time does
not change with $T$. 

In both the perception and production epochs of the interval-production
task as well as in the beating intervals in the synchronization-continuation
task, the neural network needs to sense the time flow using state
evolution along trajectories. Such state evolution may be realized
by several mechanisms. In the well-known pacemaker-accumulator model
\cite{Buhusi_2005} (recently supported in \cite{Cook_2022}), this
is realized by an accumulator that counts the number of pulses received
from a pacemaker (\textbf{Figure \ref{fig:basic_timing_model}c, left}).
In the synfire chain model \cite{Zeki_2019}, this is realized by
sequential excitation of a chain of neurons (\textbf{Figure \ref{fig:basic_timing_model}c,
middle}). In the striatal beat-frequency model \cite{Matell_2004},
this is realized by a group of oscillators with heterogeneous frequencies
whose phases are reset by the stimulus (\textbf{Figure \ref{fig:basic_timing_model}c,
right}). 

Anatomically, previous authors have identified several timing-participating
brain areas, such as basal ganglia \cite{Jin_2009}, supplementary
motor area (SMA) \cite{Mita_2009}, sensory cortex \cite{Shuler_2006},
and prefrontal cortex \cite{Wang_2018}. There is a controversy about
whether timing relies on dedicated circuits in the brain or an intrinsic
computation that emerges from the inherent dynamics of neural circuits
\cite{Paton_2018,Ivry_2008}. An influential viewpoint is that timing
depends on the interaction of core-timing areas (including basal ganglia
and SMA) that are consistently involved in temporal processing across
contexts and other areas (such as prefrontal cortex, sensory cortex
and cerebellum) that are activated in a context-dependent fashion
\cite{Merchant_2013}. 

At the behavioral level, the most well-known timing principle is the
scaling property, which states that the variance of the estimation
of time interval is proportional to the mean of the estimation \cite{Allman_2014}. 

\section*{Brain-related technology}

Time processing is also an indispensable dimension of cognition \cite{Merchant_2013};
time is also an indispensable dimension of any real-world signal to
be processed in technology. Therefore, one may expect that the results
of basic timing studies lay down foundations for processing temporal
signals in brain-related technology. This section will review two
fields of brain-related technology, neuroengineering for brain health
and brain-inspired artificial intelligence, which are two application
fields of neuroscience suggested by the China brain project \cite{Poo_2016}.
Unfortunately, we will see that the results from basic timing studies
are hardly helpful in solving practical problems.

\subsection*{Neuroengineering for brain health}

Neuroengineering refers to techniques to design interface between
living neural tissue and non-living construct, with the aim to understand,
repair, replace, or enhance neural systems \cite{Hetling_2008}. My
review on neuroengineering below will be mainly about the therapy
of Parkinson's disease by deep brain stimulation, the diagnosis of
epilepsy by neuroimaging, and the speech prosthesis by machine translating
brain activity into language. 

Parkinson's disease is closely related to the pathological change
of basal ganglia \cite{Poewe_2017}, which is a core-timing area \cite{Merchant_2013};
and epilepsy also recruits timing-related regions such as thalamus,
basal ganglia and frontal lobe \cite{Bertram_2009,Wu_2019}. As a
result, patients of either Parkinson's disease or epilepsy manifest
distortion of timing perception \cite{Gu_2016,Greyson_2014,Cainelli_2019}.
Besides, language has rich hierarchical temporal structures, and the
processing of language may share a similar neural substrate with the
processing of music \cite{Patel_2003,Janata_2003,Hickok_2012}. Therefore,
one may expect that the basic timing studies are of big help in the
therapy and diagnosis of Parkinson's disease and epilepsy, as well
as the machine translation of brain activity into language. However,
I will show below that this is not the case. 

\subsubsection*{Neuroengineering is driven by clinical data and experience}

\begin{figure}
\center \includegraphics[scale=0.6]{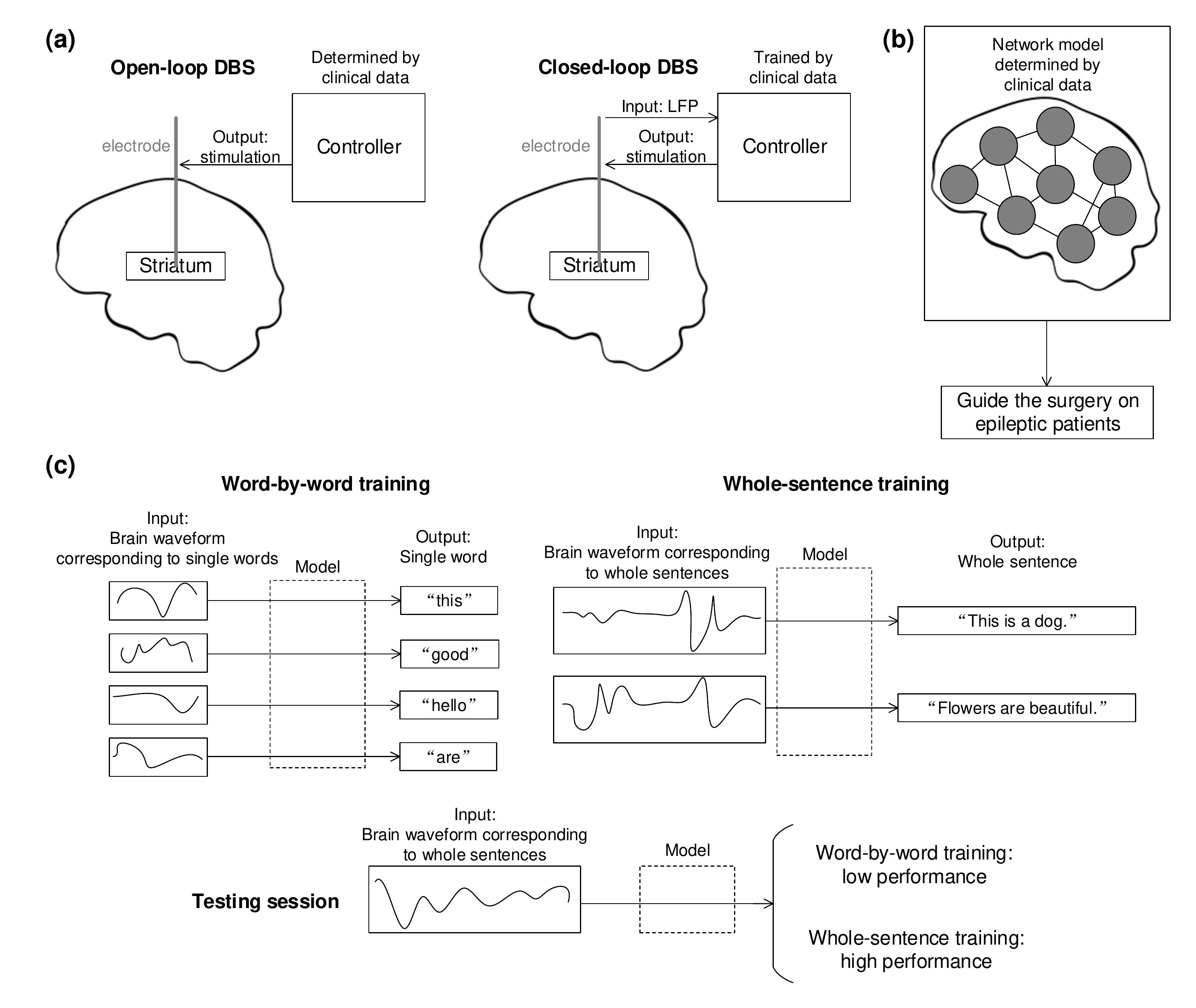}

\caption{\textbf{\label{fig:Neuroengineering-techniques}Some neuroengineering
techniques. }(\textbf{a}) In open-loop DBS (left), the stimulation
waveform is preset in the controller. In closed-loop DBS (right),
the stimulation waveform can be adjusted according to the local field
potential (LFP) of the brain. In both DBS, the controller is programmed
according to clinical data, instead of the mechanical understandings
from basic cognitive studies. (\textbf{b}) Neural network models that
simulate large-scale dynamics of the brain have been used to guide
the surgery on epileptic patients. The parameters of the model are
also determined by clinical data, instead of the mechanical understandings
from basic cognitive studies. (\textbf{c}) There are two strategies
to build the model to translate brain activities to language for speech
prosthesis. In the word-by-word training strategy (upper left), the
model is trained to map the brain waveforms when the patient is speaking
single words to the single spoken words. In the whole-sentence strategy
(upper right), the model is trained to map the brain waveforms when
the patient is speaking whole sentences to the whole spoken sentences.
In the test session (lower), the trained model is used to translate
the brain waveforms corresponding to whole sentences. The whole-sentence
training strategy results in better performance than the word-by-word
strategy. }

\end{figure}

Deep brain stimulation (DBS) therapy for Parkinson's disease was pioneered
by Lawrence Pool, who implanted an electrode into the caudate nucleus
of a female patient in 1948 \cite{Pool_1954}. Traditional DBS is
open loop, where the clinician sets parameters of the controller that
delivers short-duration (60 to 180 ms) and high-frequency (typically
130 to 185 Hz) pulses of electrical stimulation to ameliorate symptoms
\cite{Benabid_1994,Limousin_1995,Siegfried_1994} (\textbf{Figure
\ref{fig:Neuroengineering-techniques}a, left}). Such open-loop DBS
cannot adapt its stimulation according to the feedback from the patients,
and has the drawback of adverse effects such as dyskinesia as well
as high consumption of the energy in the battery \cite{Bouthour_2019,Krauss_2021}.
Recently developed closed-loop DBS overcomes these problems by delivering
stimulation only when pathological biomarkers are detected \cite{Bouthour_2019,Krauss_2021}
(\textbf{Figure \ref{fig:Neuroengineering-techniques}a, right}). 

Interestingly, despite the success and broad application of DBS, we
still do not clearly understand why DBS ameliorates Parkinson\textquoteright s
disease, although some mechanisms related to neuronal circuits, astrocytes
and neurogenesis have been proposed \cite{Okun_2012}. Due to the
not-well understanding of the mechanism, the technical details of
DBS have been mainly established empirically. For example, the optimal
stimulation waveform shape in open-loop DBS was determined by systematically
varying stimulation parameters and examining the therapeutic effects
\cite{Rizzone_2001,Kuncel_2006}. The most prominent biomarker used
in closed-loop DBS, excessively synchronized beta oscillation, was
also found by empirical comparison between normal and diseased brains
\cite{Oswal_2013,Cheyne_2013}. Therefore, mechanical insight, which
is the aim of basic timing studies (\textbf{Figure \ref{fig:basic_timing_model}}),
is not the leading force in the development of DBS. 

Though the mechanical insights provided by basic timing studies are
of little help in the current research of DBS, one may still expect
that these insights will still be helpful in the future. However,
I am not optimistic about this opinion after considering the recent
research trend: data-driven automatic design, instead of rational
implementation of mechanical knowledge, dominates the recent development
of DBS technology. As I mentioned above, closed-loop DBS delivers
stimulation into the brain only when pathological activities (i.e.,
biomarkers) are detected. Traditionally, excessive beta oscillation
was pre-assigned as the key biomarker of patients' tremors in Parkinson's
disease \cite{Bouthour_2019,Krauss_2021}. However, in two recent
studies \cite{Shah_2018,Tan_2019}, the authors recorded patients'
body movement using accelerometers and recorded local field potentials
(LFPs) using electrodes, then they trained binary classifiers to detect
the LFPs during the tremor or non-tremor periods. Here, the detector
(i.e., the binary classifier) is trained by clinical data, instead
of rationally designed using our knowledge of the mechanisms of Parkinson's
disease. A similar data-driven approach has also been used to detect
the biomarkers of depression \cite{Scangos_2021b,Scangos_2021a},
where a classifier was trained to map the stereoelectroencephalography
(SEEG) recordings to depression scores measured by a psychological
questionnaire. 

This data-driven approach is also the mainstream of other neuroengineering
techniques. For example, in two recent studies on epilepsy diagnosis,
neural network models were built to simulate the large-scale dynamics
of the brain. The models were then used to identify the ictogenic
zone of seizures and guide the resection of brain area in clinical
surgery \cite{Cao_2022,Sinha_2017} (\textbf{Figure \ref{fig:Neuroengineering-techniques}b}).
In their neural network models, the connection strengths were determined
by fitting empirical data, instead of rationally designed by some
mechanical insights into epilepsy or the information processing of
the brain. Another example is the machine translation of brain activities
to language, which can be used as speech prosthesis for degenerative
motor diseases like amyotrophic lateral sclerosis and locked-in syndrome
(\textbf{Figure \ref{fig:Neuroengineering-techniques}c}). Traditional
approaches trained translation machines by mapping neuroimaging signals
to individual words or even sub-word syllabic features (such as vowel
harmonics and fricative consonants) \cite{Pasley_2012,Angrick_2019}
(\textbf{Figure \ref{fig:Neuroengineering-techniques}c, upper left}).
But the best brain-to-language translation performance is now realized
by training recurrent neural networks in an end-to-end manner by mapping
brain signals to entire sentences rather than single words or syllabic
features \cite{Makin_2020,Cogan_2020,Moses_2019} (\textbf{Figure
\ref{fig:Neuroengineering-techniques}c, upper right}). Although understanding
the brain activities related to words and even syllables sounds more
`basic' and `mechanical', implementing such understandings in the
technique instead results in worse performance than directly training
a neural network mapping brain signals to entire sentences (\textbf{Figure
\ref{fig:Neuroengineering-techniques}c, lower}). 

\subsubsection*{Summary}

Overall, basic timing studies (\textbf{Figure \ref{fig:basic_timing_model}}),
though expected to be `basic', unfortunately do not lead the progress
of application-oriented research. This awkward situation is due to
a gap that basic timing studies aim for mechanistic explanations for
simple timing tasks, but neuroengineering, which aims for good performance
in practical use, is not mainly driven by mechanistic understandings
of the brain, but by clinical data and experience. This gap not only
exists between basic timing studies and neuroengineering, as we have
discussed here, but more generally, between basic cognitive studies
(\textbf{Figure \ref{fig:basic-cog-studies}}) and neuroengineering.
Therefore, we may conclude that basic cognitive studies do not lead
the progress of neuroengineering. 

\subsection*{Brain-inspired artificial intelligence}

Brain-inspired artificial intelligence (AI) is another potential application
field of neuroscience. Brain-inspired AI aims to build strong AI (i.e.,
AI that has mental capabilities and functions that mimic the human
brain, or in other words, can pass the Turing test) by mimicking the
structure and function of the brain, through the implementation of
neuroscience knowledge in AI engineering \cite{Hassabis_2017}. I
have a criticism of this brain-inspired approach to AI, though: due
to ethical reasons, we cannot perform detailed investigations on the
human brain, so in principle, brain-inspired AI can only closely mimic
the brain of animals, which is of low-level intelligence, instead
of the brain of humans, whose high-level intelligence is the ultimate
aim. Therefore, this brain-inspired approach should not be the leading
approach to strong AI in the long run. I will talk about the possible
approach to strong AI at the end of this subsection; at present, however,
let us forget this criticism and think about how basic timing studies
may contribute to brain-inspired AI. Unfortunately, I will show that
basic timing studies are also of little help to this field. 

\subsubsection*{The inspiration of AI from neuroscience}

The inspiration of AI from neuroscience is at the levels of neuron,
synapse, and neural network, exemplified below: 
\begin{enumerate}
\item Single neuron level
\begin{enumerate}
\item Biological neurons fire spikes, unlike artificial analog neurons,
whose activities take continuous values. Implementing spiking neurons
in hardware significantly reduces energy consumption compared with
analog neurons \cite{Frenkel_2021}. The reason is that the membrane
voltage of spiking neurons stays near the resting state most of the
time due to the sparsity of spiking periods, resulting in small leaky
currents. 
\item Biological neurons also have rich internal dynamics due to the interaction
between the membrane voltage and ion channels \cite{Dayan_2001},
unlike artificial neurons, which are usually nonlinear filters of
the total synaptic currents. Such rich internal dynamics significantly
improve the computational power of biological neurons \cite{Beniaguev_2021}.
Recently, it has been found that merely 19 neurons with internal dynamics
can make up a full-stack autonomous vehicle control system \cite{Lechner_2020}. 
\end{enumerate}
\item Single synapse level
\begin{enumerate}
\item Biological synapses have binary efficacies \cite{OConnor_2005}, unlike
artificial networks where synaptic weights usually take continuous
values. Perhaps inspired by the biological fact, binary-weight artificial
neural networks have been investigated and broadly used due to their
low computation and memory cost as well as performance comparable
with continuous-weight networks \cite{Courbariaux_2016}. 
\item Biological synapses also have hidden states other than synaptic efficacy,
due to the complex interactions of proteins in synapses \cite{Graupner_2010}.
Adding hidden synaptic states in artificial neural networks improves
memory capacity and learning performance \cite{Baldassi_2007,Kirkpatrick_2017}.
The reason is that a high hidden state of a synapse can indicate that
this synapse is important for the good performance of a task. So if
we protect the efficacy of synapses with high hidden states from being
changed in the further training process, the performance of the neural
network will not be impaired in the further training process.
\end{enumerate}
\item Neural network level
\begin{enumerate}
\item Memory replay, found in hippocampus and cortex \cite{Ji_2007}, is
a phenomenon in which the neuronal firing sequence in sleep or at
rest closely matches the firing sequence in the real experience just
before. Memory replay inspires DQN \cite{Mnih_2015}, a well-known
deep reinforcement learning algorithm that guides actions according
to perceptual inputs in order to maximize future rewards. Besides,
memory replay is also used in the Dyna algorithm \cite{Sutton_2018}
to train a mental model of the environment. After training, the agent
can predict the outcome of an action in situations never seen before
using this mental model, facilitating the agent to adapt to more complicated
environments. 
\item Biological neurons are subject to gain modulation, which means that
one input, the modulatory one, affects the sensitivity of a neuron
to another input \cite{Salinas_2000,Salinas_2001}. Gain modulation
is the neural mechanism of attention. With attention mechanism, a
neural network looks at an image or input sequence and decides which
parts of the image or sequence are important for the undertaken task,
then sends only the important parts to subsequent information processing.
Attention mechanism has been an indispensable component in advanced
image and language processing models \cite{Vaswani_2017,Devlin_2019}. 
\item Context-dependent gating \cite{Cichon_2015} means that different
sparse sets of dendritic branches are disinhibited when the brain
is involved in different tasks. Due to this mechanism, the brain recruits
different dendritic branches for different tasks, so that the synaptic
weights learned for one task will not interfere with the configuration
learned for another task. Such context-dependent gating has been implemented
in artificial neural networks to avoid catastrophic forgetting during
continual learning \cite{Manning_2020,Zeng_2019}.
\end{enumerate}
\end{enumerate}

\subsubsection*{Basic timing studies hardly inspire AI }

\begin{figure}
\center \includegraphics[scale=0.8]{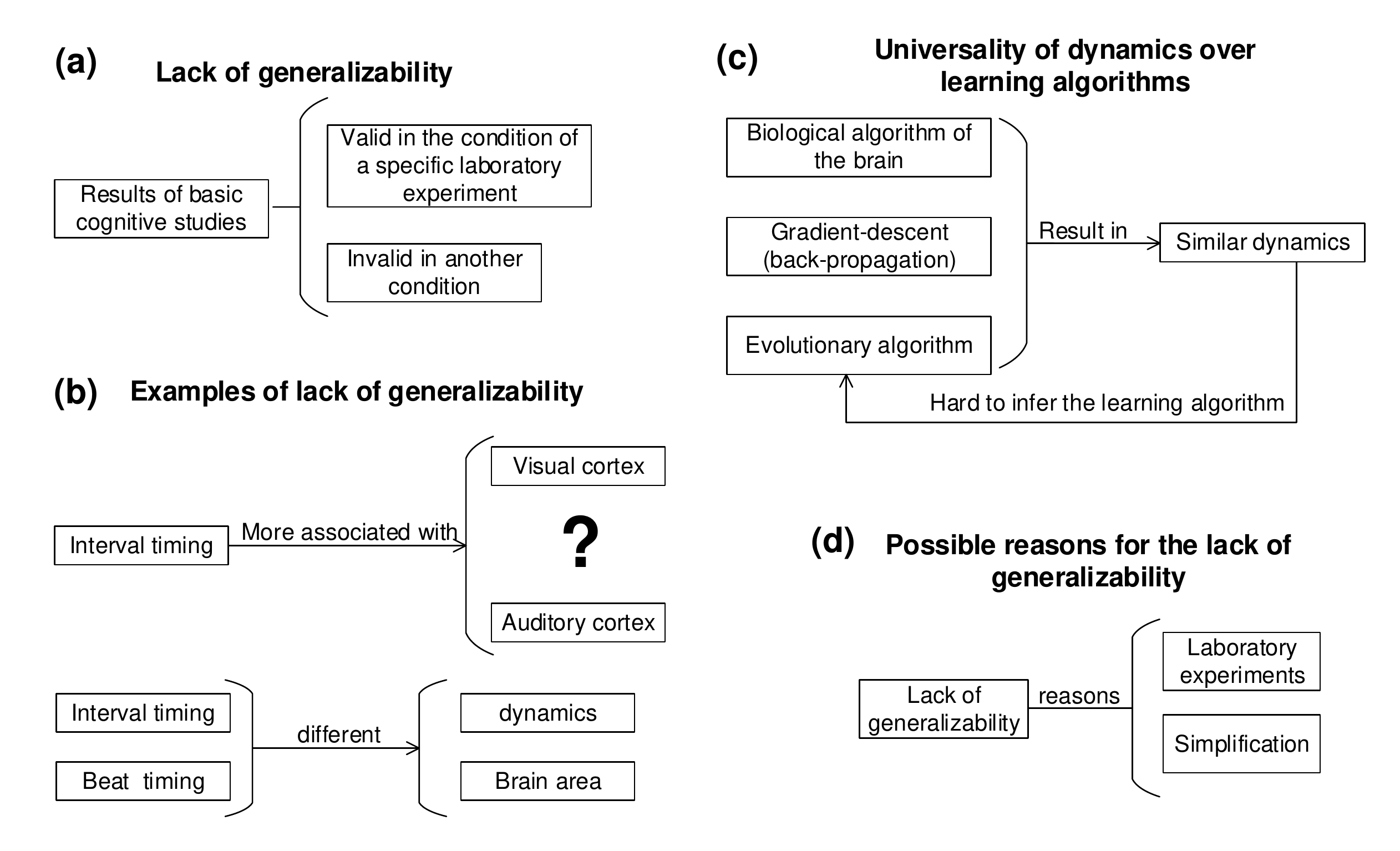}

\caption{\textbf{\label{fig:Pitfalls-of-basic-studies}Pitfalls of basic cognitive
studies.} (\textbf{a}) Basic cognitive studies may lack generalizability,
so that the results are valid in a specific laboratory experiment
condition but invalid in another condition. (\textbf{b}) Examples
of lack of generalizability. Upper: Some experiments suggest that
interval timing is more associated with visual cortex, but other experiments
suggest that interval timing is more associated with auditory cortex.
Lower: interval timing and beating timing, though both are pure timing
tasks, have different dynamic features and involve in different brain
areas. (\textbf{c}) Different learning algorithms result in similar
neuronal population dynamics when the neural network is trained on
the same basic cognitive task. Therefore, we cannot infer the learning
algorithm that the brain uses through the dynamics when the brain
is performing basic cognitive tasks. (\textbf{d}) The lack of generalizability
may be because of the methodology of laboratory experiments in basic
cognitive studies and also the misunderstanding of the ideology of
simplification. }

\end{figure}

From the examples above (also see \cite{Hassabis_2017} for a detailed
review), we see that results of basic timing studies do not go into
the mainstream of brain-inspired AI, though time processing is a fundamental
aspect of brain cognition. More generally, other basic cognitive topics,
such as working memory or decision making (\textbf{Figure \ref{fig:basic-cog-studies}b,
c}), though attract great interest in the neuroscience community,
also contribute little to brain-inspired AI. Below, I list two possible
reasons for the incapability of basic cognitive studies in AI applications:
\begin{enumerate}
\item Lack of generalizability (\textbf{Figure \ref{fig:Pitfalls-of-basic-studies}a})

All the neural mechanisms implemented in AI have a common property:
they are not task-specific. In other words, if a neural mechanism
exists only when the brain is performing a simple task like \textbf{Figure
\ref{fig:basic-cog-studies}b-e}, but does not exist if the brain
is performing another more complicated task, this neural mechanism
will not be used in AI implementation. The reason is simple: AI aims
to solve complicated real-world problems, instead of toy problems
like \textbf{Figure \ref{fig:basic-cog-studies}b-e} designed by neuroscientists. 

Lack of generalizability to real-world situations is the shortcoming
of basic cognitive studies. The dynamics of the brain when performing
complicated tasks cannot be deduced from the dynamics when the brain
is performing simple tasks. In other words, even if we have a good
understanding of the dynamics when the brain performing numerous simple
tasks like \textbf{Figure \ref{fig:basic-cog-studies}b-e}, we still
do not know the brain dynamics in complicated tasks. For example,
suppose we let a patient perform simple tasks of speaking single words.
Even if we record the brain activity related to numerous single-word
speaking, we still do not know the patient's brain activity when speaking
a whole sentence: because by dividing a sentence into single words,
we are neglecting the syntactic structure of the sentence. This is
why the translation of brain activity to language for speech prosthesis
achieves better performance when we train the neural network to translate
one sentence at a time instead of one word at a time \cite{Makin_2020,Cogan_2020,Moses_2019}
(\textbf{Figure \ref{fig:Neuroengineering-techniques}c}). More generally,
cognition requires the coordination of all `basic' elements: perception,
memory, decision making, and so on. Even if we study every `basic'
element alone well, we will still not be able to understand how the
brain performs complicated real-world tasks that require the coordination
of these elements. We will discuss more about the lack of generalizability
in the next section. 
\item Lack of insight into brain learning mechanism (\textbf{Figure \ref{fig:Pitfalls-of-basic-studies}c})

One may wonder whether the dynamic features when the brain performs
simple tasks (\textbf{Figure \ref{fig:basic_timing_model}}) depends
on the specific learning algorithm of the brain. If some learning
algorithm results in the experimentally observed features (\textbf{Figure
\ref{fig:basic_timing_model}}), whereas other algorithms do not,
we will be able to infer the brain learning algorithm through these
dynamic features. This learning mechanism, which is probably not task-specific
(therefore generalizable), can then be implemented into AI design.
Unfortunately, accumulating evidence suggests that similar dynamic
features universally emerge when neural networks are trained on the
same basic cognitive task using different learning algorithms, which
implies that we may not be able to infer critical information about
the learning algorithm that the brain uses through the dynamics when
the brain is performing simple tasks (\textbf{Figure \ref{fig:Pitfalls-of-basic-studies}c}).

For example, though error back-propagation (BP) algorithms lack convincing
experimental support \cite{Lillicrap_2020}, artificial neural networks
trained by BP exhibit biologically plausible dynamics in image classification
task \cite{Hong_2016}, language task of next-word prediction \cite{Goldstein_2022},
and other simple tasks in basic cognitive studies \cite{Bi_2020,Mante_2013}.
Recently, I trained recurrent neural networks using an evolutionary
algorithm to perform the context-dependent decision-making task \cite{Bi_2022},
and found that the resulted network exhibited dynamics closely analogous
to that observed in monkey experiments and the dynamics observed in
artificial neural networks trained by BP \cite{Mante_2013}. The reason
for this universality of dynamics over different learning algorithms
is unknown; it is possibly because different algorithms universally
tune the synaptic weights into a high entropy region in the synaptic
configuration space \cite{Baldassi_2015,Bi_2020b}. Here, `high entropy'
means that if we slightly perturb the synaptic weights found by an
algorithm, the perturbed weights still probably result in good task
performance. Therefore, the weights found by different algorithms
are likely to be close to each other in a high entropy region, which
may be the reason for the universal dynamic property of the networks
trained by different algorithms. Due to this universality, we cannot
gain insight into the learning mechanism of the brain from the dynamic
features found in basic cognitive studies, not to mention to implement
the brain learning mechanism in AI. 
\end{enumerate}

So what is the approach to strong AI, the machine with intelligence
equal to the human brain or even more powerful? In my opinion, the
most important thing we should learn from biology is the colossal
scale of the human brain. Comparative studies suggest that human brain
contains more neurons than any other animal, which is probably the
reason for our superior cognitive abilities \cite{Herculano-Houzel_2012}.
Consistently, AI is undergoing a paradigm shift with the rise of colossal
models (e.g., BERT \cite{Devlin_2019} and GPT-3 \cite{Brown_2020})
with over 100 billion parameters trained on oceans of data \cite{Bommasani_2021}.
Such models, trained unsupervisedly, develop geometric representation
of knowledge \cite{Manning_2020,Rives_2021}, which versatilely serve
as the common basis of many task-specific models via adaptation \cite{Bommasani_2021}.
Most impressively, as the size of the neural network increases, advanced
functionalities such as in-context learning naturally emerge \cite{Brown_2020}:
in-context learning means that the neural network after training can
be competent for a task never seen during training, after the trained
neural network is instructed by a natural-language description of
the task. Such colossal models are becoming the trend of AI led by
big tech companies such as Google, Microsoft and Huawei, with broad
applications in text \cite{Devlin_2019}, images \cite{Ramesh_2021},
protein design \cite{Rives_2021}, and chemical reactions \cite{Schwaller_2021}.
Recently, at a conference, a manager of the colossal-model project
of Huawei told me that the progress of colossal models is also gradually
getting stagnant, because we cannot afford the huge energy consumption
to train the model if the model is too large \cite{Strubell_2019}.
He believed that the next generation of colossal models should be
led by the revolution of computation paradigm, such as quantum computing,
which can speed up some kinds of computations by exponential order
\cite{Nielsen_2011}. At another conference, a professor of quantum
physics told me that quantum algorithms are getting mature; the bottleneck
of quantum computing lies in its hardware implementation. If their
opinions are correct, we may expect that strong AI will naturally
emerge after the manufacturing technology of quantum computers is
mature and if we train mega-colossal models using quantum computers. 

\subsubsection*{Summary}

Overall, basic cognitive studies cannot significantly contribute to
AI due to its limitations on generalizability and also the insight
into brain learning mechanism. The future of AI is likely to be led
by colossal models.

\section*{Contemplation}

Time processing is an indispensable dimension of cognition, and time
is also an indispensable dimension of any real-world signal to be
processed in technology. But why do basic timing studies, which aim
to study time processing in the brain, which are also interesting
and elegant, instead have little help to the progress of brain-related
technology? Below, I will discuss possible reasons. 

\subsection*{Generalizability---the shortcoming}

Generalizability is a measure of how useful the results of a study
are for broader situations. Generalizability is the critical hypothesis
(and also the aim) of science. To understand this point, let us consider
a simple example. If we want to test the effectiveness of a new drug,
we will recruit several patients and test the drug on them. However,
our aim is not only to investigate these several recruited patients,
but to draw a general conclusion on the effect of the drug on the
mass of people using these recruited patients, with the hypothesis
that similar phenomena can also be observed if we recruit another
group of patients. As another example, when physicists perform an
experiment and conclude a physical law, their aim is not only to explain
the very experiment they perform, but to conclude a law generalizably
applicable to other experiments taken at another place and another
time. Therefore, generalizability is also the hypothesis underlying
their research. However, we should not take such generalizability
for granted. Many hard problems are because we do not have a generalizable
understanding of the problem or a generalizable technique to deal
with the problem. For example, cancer is hard to cure because we do
not have a generalizable technique to efficiently kill all the cancer
cells due to the high diversity of cancer cells \cite{Morita_2020,Black_2021}. 

Lack of generalizability is a shortcoming of basic timing studies:
the result obtained under one experimental condition often cannot
predict the result under another condition. For example, suppose an
auditory and an visual stimuli are respectively associated with two
different time durations of $T_{a}$ and $T_{v}$ in a rat subject.
In that case, the presentation of the auditory and visual stimuli
simultaneously will make the rat subject to time an expected duration
of $T_{+}$, with $T_{+}$ between $T_{a}$ and $T_{v}$ but closer
to $T_{v}$ \cite{Swanton_2011,Matell_2014}. Additionally, compared
with an auditory stimulus, the association between a visual stimulus
with a time duration can be better transferred to a subsequent operant
response when tested in a Pavlovian-instrumental transfer procedure
\cite{Matell_2017}. These results imply that visual signals are more
involved in interval timing than auditory signals. However, in a recent
study on an action timing task, in which a mouse had to learn the
timing of its action based on the sensory feedback caused by the action
of itself, it was the deprivation of auditory (not visual) input that
disrupted the learned action timing \cite{Cook_2022}, which contradicts
with the understanding before (\textbf{Figure \ref{fig:Pitfalls-of-basic-studies}b,
upper}). As another example, there are two frequently studied experimental
paradigms of timing tasks: one is interval timing, in which the subject
is to perceive or produce a single time interval \cite{Karmarkar_2007,Wang_2018}
(\textbf{Figure \ref{fig:basic-cog-studies}d}); the other is beating
timing, in which the subject is to perceive or produce regular beats
\cite{Gamez_2019} (\textbf{Figuer \ref{fig:basic-cog-studies}e}).
It has been found that the brain uses different neural substrates
and mechanisms to process temporal information in these two paradigms
\cite{Teki_2011,Karmarkar_2007,Wang_2018,Gamez_2019}, even though
they are both pure timing tasks with no other information (such as
spatial information) involved (\textbf{Figure \ref{fig:Pitfalls-of-basic-studies}b,
lower}). Overall, the lack of generalizability makes us hard to conclude
how the brain processes temporal information. 

Below, I discuss two possible reasons underlying the lack of generalizability
(\textbf{Figure \ref{fig:Pitfalls-of-basic-studies}d}):
\begin{enumerate}
\item Laboratory experiments

Basic timing studies are performed in laboratory experiments, whose
experimental conditions are artificially designed and well-controlled
(just like \textbf{Figure \ref{fig:basic-cog-studies}d, e}) instead
of in real life. The lack of generalizability has long been recognized
as the shortcoming of laboratory experiments in social science (including
psychology) \cite{Bruggemann_2016,Hulstijn_1997}, so the problems
of basic timing studies discussed here are only examples of the general
shortcoming of the paradigm of laboratory experiments. 
\item Misunderstanding of simplification

Simplification is a pervasive idea in the data analysis and computational
models of basic timing studies (\textbf{Figure \ref{fig:basic_timing_model}}).
The best-known timing model, the pacemaker-accumulator model \cite{Buhusi_2005}
(\textbf{Figure \ref{fig:basic_timing_model}c, left}), contains only
four components (pacemaker, accumulator, memory device and comparator)
to model the timing process. The dynamic features found by basic timing
studies (\textbf{Figure \ref{fig:basic_timing_model}a, b}) are often
discovered after reducing the dimension of the population dynamics
of neural networks using principal component analysis (PCA). This
PCA method also manifests the idea of simplification: simplifying
the population dynamics by reducing its dimension. 

The idea of simplification, also named the principle of Occam's razor,
tries to explain the world using as few entities as possible. However,
we may have to understand the advantage of this principle before using
it. A widely accepted advantage of simplification is that simple theories
tend to be more testable and, therefore, easier to be falsified \cite{Baker_2016,Sober_1991}.
In other words, the main advantage of a simple theory is not that
it can better predict the experiment, but instead lies in its easiness
of falsification, which is believed to be a necessary property of
a scientific theory \cite{Popper_1959}. Another advantage of simplification
(with controversy) is that it improves induction: choosing a simple
theory after numerous observations reduces the change of theory after
more future observations \cite{Baker_2016}. This induction advantage
is closely related to the concept of generalizability we discuss here,
because reducing the change of theory after future observations means
improving the generalizability of the theory. However, `induction'
means that the theory must be concluded after numerous observations,
which is apparently not the case for the results (\textbf{Figure \ref{fig:basic_timing_model}})
in basic timing studies, which are usually proposed based on single
laboratory experiments under simple and well-controlled situations.
In other words, if we indeed want a simple timing theory generalizable
to real-world situations, numerous observations in real-world situations
are necessary. 
\end{enumerate}

\subsection*{Unknowability---the reality }

\begin{figure}
\center \includegraphics[scale=0.9]{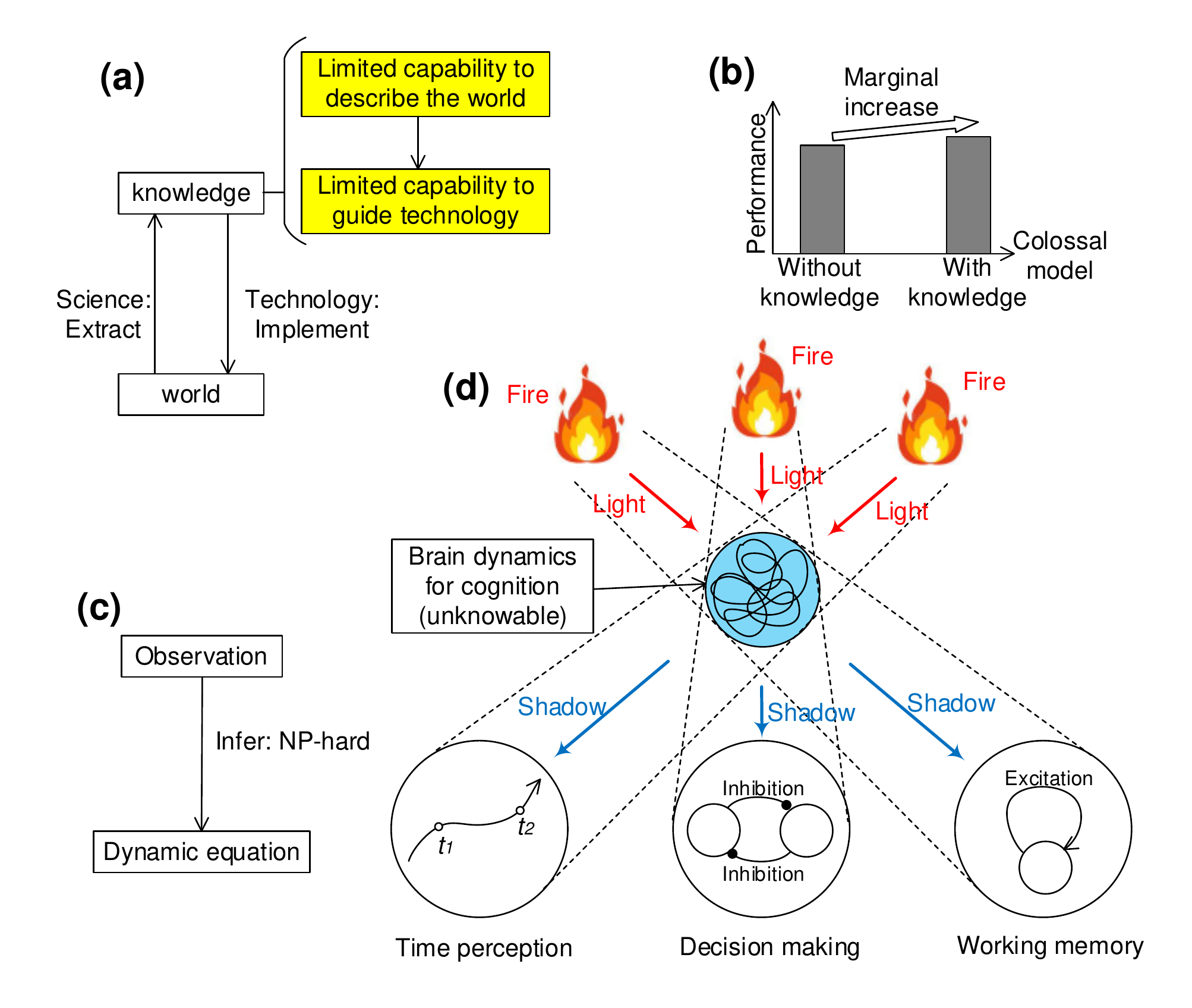}

\caption{\textbf{\label{fig:unknowability}The unknowability of the world.}
(\textbf{a}) Science extracts knowledge from the world, and technology
implements knowledge to change the world. However, if knowledge has
limited capability to describe the world, it will also have limited
capability to guide the creation of technology (yellow boxes). (\textbf{b})
After adding knowledge into a colossal model, its performance only
marginally improves. (\textbf{c}) Inferring the dynamic equation from
experimental observation is an NP-hard problem. (\textbf{d}) A Platonic
understanding of cognition. The dynamics of the brain in various tasks
(such as the stereotypical trajectory for time perception \cite{Bi_2020,Karmarkar_2007},
competing dynamics by mutual inhibition for decision making \cite{Wong_2006},
and self-excitation for working memory \cite{Lim_2013}) are various
shadows of an unknowable object (the brain dynamics for cognition)
under fires at different positions. This Platonic viewpoint implies
that we will still not understand cognition after studying the dynamics
in various tasks, unlike the atomic viewpoint (\textbf{Figure \ref{fig:basic-cog-studies}a}),
which believes that we will understand cognition after studying each
element of cognition. }

\end{figure}

Basic cognitive studies record the brain activity when subjects are
performing deliberately designed tasks (\textbf{Figure \ref{fig:basic-cog-studies}b-e}),
aiming to draw knowledge of the neural mechanisms of cognition. This
methodology is based on the following philosophical understanding
of science (\textbf{Figure \ref{fig:unknowability}a}): science investigates
the world, generates knowledge, then technology uses the knowledge
generated by science to change the world. However, the pitfall of
this philosophy is that it does not consider the possibility that
the capability of knowledge (therefore, science) to describe the world
is fundamentally limited. In other words, some parts of the world
are unknowable. If this is the case, the knowledge generated by science
will not be able to well guide the design of technology to change
the world (\textbf{Figure \ref{fig:unknowability}a}). The lack of
generalizability discussed before may also stem from the unknowability
of the world (including cognition): if the capability of knowledge
to describe the world is fundamentally limited, we should not dream
of the luxury that our knowledge has the possibility to generalize
to every situation. 

\subsubsection*{The inspiration from AI}

To discuss the limited capability of knowledge, let us start with
an interesting empirical finding in AI. In AI, knowledge is usually
represented by \textit{(subject; relation; object)} triplets, representing
the relationship between a subject and an object \cite{Hogan_2021}.
For example, the sentence `dog is animal' can be represented by a
triplet \textit{(dog; is; animal)}. A collection of a large number
of triplets is called a knowledge base. It has been found that adding
knowledge bases to deep learning models can improve the performance
of natural language processing \cite{Guo_2022,Annervaz_2018}. However,
interestingly, well-known colossal models (such as GPT-3 \cite{Brown_2020}
of Microsoft or Pangu \cite{Zeng_2021} of Huawei) are pure deep neural
networks without a knowledge base. A possible explanation for why
well-known colossal models do not contain a knowledge base is that
the performance improvement after adding a knowledge base to colossal
models is marginal (below 4\%, see Table 5 of \cite{Colon-Hernandez_2021})
(\textbf{Figure \ref{fig:unknowability}b}). I have discussed this
interesting phenomenon with an AI expert in NetEase, who believed
that this is because colossal models are trained by oceans of texts
collected from the internet, which contains far richer information
than knowledge bases can provide, so adding knowledge bases into colossal
models can hardly increase the information used to train the colossal
models. Notice that people have invested great efforts to develop
knowledge bases: for example, well-known knowledge bases such as YAGO
and Freebase contain over 1 billion triplets. Despite such efforts,
these knowledge bases are still hardly useful in the core AI technology
of colossal models. 

What does this empirical finding of AI tell us? Notice that science
is a process of generating knowledge from experiments (\textbf{Figure
\ref{fig:unknowability}a}): for example, basic timing studies aim
to establish the relation between the dynamics of the brain and the
behavioral task. Also, notice that AI represents the future of technology.
Therefore, if knowledge bases cannot help AI, we may conclude that
science will not help technology in the future! 

\subsubsection*{The inspiration from philosophy and physics}

The recognization of the limited capability of knowledge has a long
history in philosophy. David Hume believed that causality cannot be
justified, because we can only observe that one thing $A$ happened
after another thing $B$, but cannot observe the underlying causal
mechanism that made $A$ happen after $B$ \cite{David_Hume}. Immanuel
Kant believed that there exist things (the so-called `things-in-themselves')
that are unperceivable and unknowable, what we can perceive are mere
`appearances' of these unknowable things, and a theory of the world
develops when the perceived things conform to our spatial and temporal
forms of intuition \cite{Immanuel_Kant}. In 1963, Frederic Fitch
proposed a logic paradox, which asserts that if all truths were knowable,
it would follow that all truths are already known \cite{wiki_knowability}.
Therefore, if we acknowledge that not all truths are already known,
we have to acknowledge that not all truths are knowable. Fitch's paradox
sets up a fundamental limitation of the capability of experiments:
there exists truth that cannot be known using experiments, no matter
how advanced techniques we use. 

A recent physical study further explains the idea of unknowability.
That study shows that identifying the underlying dynamical equation
(i.e., the physical reality) from any amount of experimental observations
is provably NP-hard both for classical and quantum mechanical systems
\cite{Cubitt_2012a,Cubitt_2012b} (\textbf{Figure \ref{fig:unknowability}c}).
In other words, if $\text{NP}\neq\text{P}$, which is belived by most
computer scientists, identifying the underlying dynamical equation
will take us time with duration exponential to the dimension of the
system, so that the dynamical equation will be essentially unknowable
if the dimension of the system is large. 

There have been enormous studies about NP problems using models (e.g.,
spin glass) derived from statistical physics \cite{Mezard_2009},
which gain insight into the nature of the computational difficulty
in solving these problems. The main result is that the computational
difficulty is closely related to the (some kind of ) correlation between
degrees of freedom in the system. To briefly understand the idea,
let us consider a system that can be described by a state $\boldsymbol{x}=(x_{1},x_{2},\cdots,x_{n})$.
If different $x_{i}$s ($i=1,2,\cdots,n$) do not interact each other,
we can find the optimal state $\boldsymbol{x}_{\text{opt}}$ of the
system with respect to a problem by sequentially optimizing each $x_{i}$
respectively. But if different $x_{i}$s strongly interact with each
other, we may have to adjust a large number of $x_{i}$s simultaneously
during the optimization process, increasing the difficulty in finding
$\boldsymbol{x}_{\text{opt}}$. 

Basic cognitive studies (\textbf{Figure \ref{fig:basic-cog-studies}})
aim to understand the dynamics of the brain underlying cognition by
observing the brain dynamics when the brain is performing simple tasks.
Therefore, basic cognitive studies address the same type of NP-hard
problem studied in \cite{Cubitt_2012a,Cubitt_2012b} that infers dynamics
from observation. We have mentioned in the last paragraph that the
difficulty of this problem lies in the correlation between different
degrees of freedom. Therefore, the atomic philosophy (\textbf{Figure
\ref{fig:basic-cog-studies}a}), which aims to understand cognition
by studying each individual cognitive element (such as perception,
memory, decision making, etc.), is actually unsuitable for guiding
cognition research. The reason is that the coordination between different
cognitive elements is important for performing a real-life task, so
it is important to consider the whole thing simultaneously. We have
mentioned a good example before (\textbf{Figure \ref{fig:Neuroengineering-techniques}c}):
the translation of brain activity to language for speech prosthesis
achieves better performance when we train the neural network to translate
one sentence at a time instead of one individual word at a time \cite{Makin_2020,Cogan_2020,Moses_2019}. 

Unfortunately, atomism is just the very philosophy that guides basic
cognitive studies (including basic timing studies), which is possibly
the reason for the difficulty we encounter in understanding cognition.
Despite decades of studies, we still do not completely understand
the processing of time in the brain: the results of basic timing studies
sometimes contradict each other (\textbf{Figure \ref{fig:Pitfalls-of-basic-studies}b})
and cannot guide technology design. Another well-known example is
the study of hippocampus. It has been found that hippocampus transfers
memory into cortex \cite{Goto_2021} and performs inferential reasoning
\cite{Barron_2020}; hippocampus encodes place \cite{Sosa_2021},
head directions \cite{Sosa_2021}, time \cite{Eichenbaum_2014}, visual
and auditory stimuli \cite{Goto_2021,Turk-Browne_2019}, and abstract
knowledge \cite{Nieh_2021}. However, up to now, we still do not clearly
understand the functional role of hippocampus; in other words, we
cannot predict the functional role of hippocampus in a new experimental
condition. What is the mechanism of the brain to process time? What
is the functional role of hippocampus? Perhaps they are essentially
unknowable. 

So how to understand the kaleidoscopic observations in timing and
hippocampal studies? In the famous allegory of the cave, Plato likens
our understanding of the world to the shadows on the wall of a cave,
cast by objects in front of a fire \cite{wiki_cave}. Inspired by
him, I think the best way to understand the observations in timing
or hippocampal studies is to regard the brain dynamics in different
experimental conditions as the shadows cast by an object from fires
at different positions (\textbf{Figure \ref{fig:unknowability}d}).
The object represents the reality of the neural mechanism of cognition,
which is unknowable, but what we can observe is only the dynamics
of the brain when performing a specific task. When the fire is at
different positions, the projection on the wall is different, similar
to the kaleidoscopic dynamics of the brain when performing different
tasks. This Platonic viewpoint implies that we may still not understand
cognition after studying the dynamics in various tasks, unlike the
atomic viewpoint (\textbf{Figure \ref{fig:basic-cog-studies}a}),
which believes that we will understand cognition after studying each
element of cognition. Plato encouraged us to walk out of the cave
and know the reality of the world through reason. But unfortunately,
inferring the reality from observation is NP-hard \cite{Cubitt_2012a,Cubitt_2012b},
so the reality may essentially be unknowable. 

\subsection*{Summary}

The reason why the results of basic cognitive studies have little
help to the progress of brain-related technology is because of their
lack of generalizability. This lack of generalizability may be rooted
in the unknowability of cognition. 

\section*{Outlook}

What can we learn from the understandings above to guide our future
research? I give three suggestions, explained in three subsections
below (\textbf{Figure \ref{fig:suggestions}}).

\subsection*{Improving generalizability }

\begin{figure}
\center \includegraphics{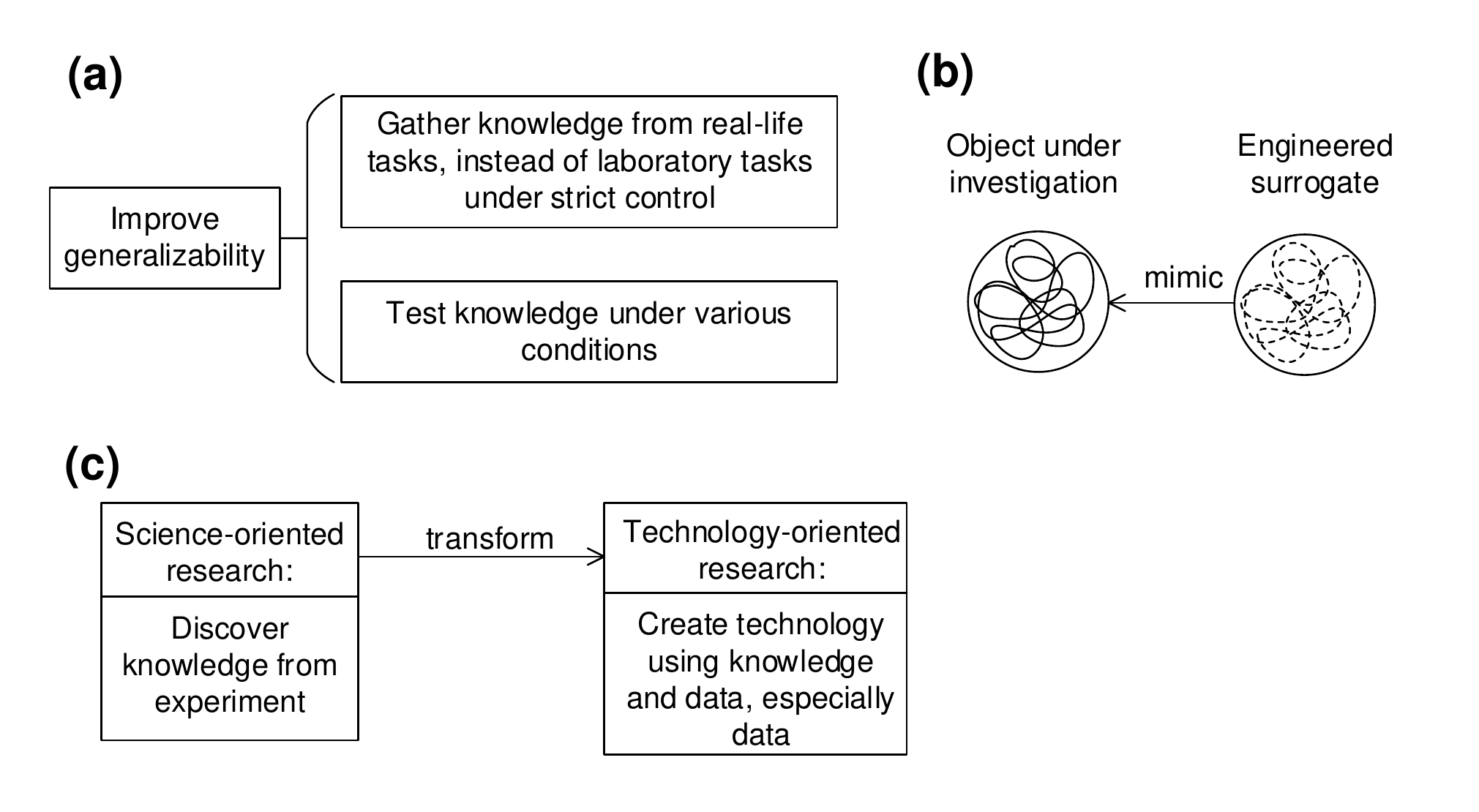}

\caption{\textbf{\label{fig:suggestions}My suggestions for future research.}
(\textbf{a}) Improve generalizability. (\textbf{b}) Engineering an
easily accessible surrogate to mimic the object under investigation,
so that we can easily predict the behavior of the object by investigating
the surrogate without experimenting on the object. (\textbf{c}) Our
research style should transform from science-oriented (which aims
to discover knowledge) to technology-oriented (which aims to create
technology using knowledge and, especially, data).}
\end{figure}

As I mentioned above, generalizability is a central aim of science.
We want that our results are valid in broader conditions, instead
of only in the condition we investigated (\textbf{Figure \ref{fig:suggestions}a}).
Basic cognitive studies (\textbf{Figure \ref{fig:basic-cog-studies}b-e})
are performed in laboratory experiments, where experimental conditions
are artificially designed and well-controlled instead of in real life.
As I mentioned before, lack of generalizability has long been recognized
as a shortcoming of laboratory experiments \cite{Bruggemann_2016,Hulstijn_1997}.
Therefore, a possible way to improve the generalizability of our results
is to extract the features of brain dynamics when the subjects are
performing real-life tasks, instead of the tasks deliberately designed
in experiments. Besides, it is necessary to verify a result under
various conditions to improve the generalizability of the result. 

\subsection*{Making use of surrogate}

The unknowability of the world implies that it is impossible to develop
a universal theory generalizable to every situation. How should we
perform research in this case? An emerging methodology to overcome
the unknowability is to use surrogates. This means that we do not
make efforts to know the mechanism underlying the phenomena of a complex
system, but instead engineer a surrogate of the system, so that the
surrogate behaves similarly to the original system in situations of
interest. In this way, we can predict the behavior of the original
system in a new situation by observing the behavior of the surrogate,
which is often more accessible than investigating the original system
(\textbf{Figure \ref{fig:suggestions}b}). Interestingly, this engineering
can be realized even if we do not know the mechanism of how the original
system works. 

Surrogating is the exact idea of neural network models. After we know
the signal $I_{i}$ ($i=1,2,3,\cdots$) inputted to a system and the
output $O_{i}$ of the system in response to $I_{i}$, we can build
up a deep neural network model by training the neural network to output
$O_{i}$ when being inputted $I_{i}$. In this way, the deep network
surrogates this system, and can be used to predict the output of the
investigated system in response to a new input signal $I'$, as long
as $I'$ is not far different from the set $\{I_{i}\}_{i}$. In this
methodology, we do not deliberately study the response mechanism of
the original system; instead, such knowledge is hidden in the parameters
of the neural network after training. Though this knowledge cannot
be understood, it can be used by us. This kind of knowledge is called
dark knowledge \cite{Jia_2019,Hinton_2015}, in comparison with light
knowledge, which is the knowledge that can be expressed by language
and formula. 

Surrogating has been widely used in brain research. It has been found
that artificial neural networks, after training on a task, exhibit
similar dynamics to the brain when performing the same task \cite{Hong_2016,Goldstein_2022,Mante_2013},
so artificial neural networks can be used as surrogates to study the
brain: this idea has been used to study timing tasks \cite{Bi_2020}.
Besides, as I mentioned before, neural network models have also been
used as surrogates for epileptic brains and guide clinical surgery
\cite{Cao_2022,Sinha_2017}. 

Moreover, surrogating has also been widely used in other areas than
brain research. Organoids, which are self-organized 3-dimensional
tissue cultures derived from stem cells, have been used in resembling
and studying various organs, modeling personalized disease, and discovering
drugs \cite{Chiaradia_2020,Kim_2020}. Digital twins, which are virtual
models designed to accurately reflect physical objects and get updated
from real-time data, have been used in designing, manufacturing, monitoring
and diagnosing large equipment such as bridges, aircrafts, and power
generators \cite{Liu_2021}. All these examples reflect the idea of
using a surrogate easier to investigate and manipulate than the original
system to study the original system, though the surrogate itself may
also be too complex to be ultimately knowable (e.g., the dark knowledge
in artificial neural networks). 

\subsection*{Being technology-oriented}

Science is the process of exploring new knowledge through observation
and experiments. Technology is the process of applying scientific
knowledge for various purposes. However, the unknowability of the
world sets up a fundamental limitation of the capability of science
to know the world and, therefore, to guide technology (\textbf{Figure
\ref{fig:unknowability}a}). Therefore, in my opinion, future studies
are going to be technology-oriented (\textbf{Figure \ref{fig:suggestions}c}),
which has the following two meanings:
\begin{enumerate}
\item Instead of being driven by the interest in how nature works, scientists
should perform their research with a blueprint (at least a sketch)
in their minds, a blueprint of how their findings may be applied to
solve practical problems. Without the guidance of technology, scientific
results tend to be incapable of guiding practical applications, just
as basic timing studies are useless in neuroengieering and brain-inspired
AI. 
\item Technology tends to be created without the guidance of scientific
knowledge. There is a trend to create technology by human-guided self-organization,
rather than by implementing knowledge through the rational design
of human. The decay of rational design is perhaps because of the unknowability
of the world. As a result, knowledge becomes increasingly useless
in dealing with complex problems. Self-organization is a process where
collective order arises from local interactions between parts of an
initially disordered system \cite{wiki_Self_organization,wiki_Spontaneous_order}.
The training of deep artificial neural networks is a self-organization
process under human guidance: we only adjust the interactions between
artificial neurons by adjusting the synaptic weights instead of directly
designing the activity of each neuron, but the collective dynamics
of the neural network when performing tasks emerges from these interactions.
A good example of the paradigm shift of technology from rational design
to human-guided self-organization is natural language processing.
The traditional method to translate one language to another was to
recognize the human-designed grammatic structure of an input sentence
and then translate the sentence based on the grammatic structure using
human-designed rules \cite{Cambria_2014}. Today, however, language
translation is based on end-to-end training of neural networks, with
the grammatic structure and translation rules automatically and implicitly
emerging during training \cite{Goldberg_2017}. As I mentioned before,
such automatic and implicit feature extraction by neural networks
has also been used to recognize pathological biomarkers in closed-loop
deep brain stimulation \cite{Scangos_2021b,Scangos_2021a} and translate
brain activity to natural language \cite{Makin_2020,Cogan_2020} (\textbf{Figure
\ref{fig:Neuroengineering-techniques}}). 
\end{enumerate}
So how to guide the self-organization of a complex system to create
technology? The dominating methodology, the deep learning technique,
is to adjust the synaptic weights of a deep network by gradient-based
algorithms while fixing the network architecture at the form pre-assigned
by human \cite{Goodfellow_2016}. Evolutionary algorithms can adjust
not only synaptic weights but also network architecture \cite{Stanley_2019}.
Therefore, compared with gradient-based algorithms, evolutionary algorithms
do not require human to design network architecture. Therefore, in
a neural network created by evolutionary algorithms, everything (either
synaptic weights or network architecture) emerges from self-organization,
and nothing is implemented by the rational design of human: this minimizes
the interference of rational design, whose capability is limited due
to the unknowability of the world, potentially leading to superior
technology \cite{Stanley_2015}. Besides, human-guided evolution is
not only an algorithm that runs in computers but also a practice in
labs. We create high-yield plants and animals by selecting their breeding
\cite{wiki_selective_breeding}; we also discover drugs and functional
proteins by directing the evolution of engineered microbes \cite{Davis_2017,Romero_2009}.
Human-guided evolution, without the need for rational design, may
be the ultimate method to create something to our desired end in this
unknowable world. 

\section*{Conclusion}

In this paper, I review the main results of basic timing studies,
and manifest their little help in solving practical problems in the
fields of neuroengineering and brain-inspired AI. Basic timing studies
extract knowledge from deliberately designed simple tasks. In contrast,
neuroengineering is mainly driven by clinical data, and AI is mainly
driven by training colossal models using oceans of data collected
from the internet. The limitation of basic timing studies is perhaps
because of the lack of generalizability of the results of basic timing
studies, which are rooted in the fundamental unknowability of the
world (including cognition). The reason for this limitation is also
true for, more generally, basic cognitive studies. Therefore, I question
and criticize the usefulness and prospect of the research protocol
of basic cognitive studies (\textbf{Figure \ref{fig:basic-cog-studies}}):
recording the brain activity when the subject is performing deliberately
designed experiments, with the aim of understanding the neural mechanism
of cognition. I then give three suggestions for future research: improving
the generalizability of results, considering using surrogates to overcome
the unknowability, and performing technology-oriented studies.

The neuroscience problem I point out here belongs to a trend of biology,
which gradually relies on mass-scale technology (such as multi-omic
databases and super-computing power) to solve practical problems using
AI \cite{Subramanian_2020}. The knowledge required by AI to solve
practical problems is not implemented by human through rational design,
but instead emerges self-organizedly in the problem-solving process
in a hidden manner. Such knowledge is encoded in the AI system (e.g.,
in the synaptic weights), but unknowable by human. We can imagine
that in the far future, when AI becomes far more powerful than human
intelligence, we may feel hard to understand the logic of AI in solving
a problem even if AI tries hard to explain it to us. Therefore, the
use of hidden knowledge, something we can use but not understand,
should be a gradually dominating paradigm in scientific and technological
research. 

\section*{Acknowledgment}

Z.B. was supported by the National Natural Science Foundation of China
(32000694).

\bibliographystyle{nature/naturemag}
\bibliography{reference}

\begin{thebibliography}{100}
\expandafter\ifx\csname url\endcsname\relax
  \def\url#1{\texttt{#1}}\fi
\expandafter\ifx\csname urlprefix\endcsname\relax\def\urlprefix{URL }\fi
\providecommand{\bibinfo}[2]{#2}
\providecommand{\eprint}[2][]{\url{#2}}

\bibitem{Feynman_2011}
\bibinfo{author}{Feynman, R.~P.}, \bibinfo{author}{Leighton, R.~B.} \&
  \bibinfo{author}{Sands, M.}
\newblock \emph{\bibinfo{title}{The Feynman Lectures on Physics, vol 1}}
  (\bibinfo{publisher}{Basic Books; New Millennium ed. edition},
  \bibinfo{year}{2011}).

\bibitem{Baldwin_1893}
\bibinfo{author}{Baldwin, J.~M.}
\newblock \emph{\bibinfo{title}{Elements of Psychology}}
  (\bibinfo{publisher}{Macmillan and Co.}, \bibinfo{address}{London},
  \bibinfo{year}{1893}).

\bibitem{Constantinidis_2001}
\bibinfo{author}{Constantinidis, C.}, \bibinfo{author}{Franowicz, M.~N.} \&
  \bibinfo{author}{Goldman-Rakic, P.~S.}
\newblock \bibinfo{title}{Coding specificity in cortical microcircuits: A
  multiple-electrode analysis of primate prefrontal cortex}.
\newblock \emph{\bibinfo{journal}{J. Neurosci.}} \textbf{\bibinfo{volume}{21}},
  \bibinfo{pages}{3646--3655} (\bibinfo{year}{2001}).

\bibitem{Roitman_2002}
\bibinfo{author}{Roitman, J.~D.} \& \bibinfo{author}{Shadlen, M.~N.}
\newblock \bibinfo{title}{Response of neurons in the lateral intraparietal area
  during a combined visual discrimination reaction time task}.
\newblock \emph{\bibinfo{journal}{J. Neurosci.}} \textbf{\bibinfo{volume}{22}},
  \bibinfo{pages}{9475--9489} (\bibinfo{year}{2002}).

\bibitem{Rakitin_1998}
\bibinfo{author}{Rakitin, B.~C.}, \bibinfo{author}{Gibbon, J.},
  \bibinfo{author}{Penney, T.~B.} \& \bibinfo{author}{Malapani, C.}
\newblock \bibinfo{title}{Scalar expectancy theory and peak-interval timing in
  humans}.
\newblock \emph{\bibinfo{journal}{J. Exp. Psychol. Anim. Behav. Process}}
  \textbf{\bibinfo{volume}{24}}, \bibinfo{pages}{15--33}
  (\bibinfo{year}{1998}).

\bibitem{Merchant_2013}
\bibinfo{author}{Merchant, H.}, \bibinfo{author}{Harrington, D.~L.} \&
  \bibinfo{author}{Meck, W.~H.}
\newblock \bibinfo{title}{Neural basis of the perception and estimation of
  time}.
\newblock \emph{\bibinfo{journal}{Annu. Rev. Neurosci.}}
  \textbf{\bibinfo{volume}{36}}, \bibinfo{pages}{313--336}
  (\bibinfo{year}{2013}).

\bibitem{Poo_2016}
\bibinfo{author}{m.~Poo, M.} \emph{et~al.}
\newblock \bibinfo{title}{China brain project: Basic neuroscience, brain
  diseases, and brain-inspired computing}.
\newblock \emph{\bibinfo{journal}{Neuron}} \textbf{\bibinfo{volume}{92}},
  \bibinfo{pages}{591--596} (\bibinfo{year}{2016}).

\bibitem{Gamez_2019}
\bibinfo{author}{G{\'a}mez, J.}, \bibinfo{author}{Mendoza, G.},
  \bibinfo{author}{Prado, L.}, \bibinfo{author}{Betancourt, A.} \&
  \bibinfo{author}{Merchant, H.}
\newblock \bibinfo{title}{The amplitude in periodic neural state trajectories
  underlies the tempo of rhythmic tapping}.
\newblock \emph{\bibinfo{journal}{PLoS Biol.}} \textbf{\bibinfo{volume}{17}},
  \bibinfo{pages}{e3000054} (\bibinfo{year}{2019}).

\bibitem{Bi_2020}
\bibinfo{author}{Bi, Z.} \& \bibinfo{author}{Zhou, C.}
\newblock \bibinfo{title}{Understanding the computation of time using neural
  network models}.
\newblock \emph{\bibinfo{journal}{Proc. Natl. Acad. Sci. USA}}
  \textbf{\bibinfo{volume}{117}}, \bibinfo{pages}{10530--10540}
  (\bibinfo{year}{2020}).

\bibitem{Jin_2009}
\bibinfo{author}{Jin, D.~Z.}, \bibinfo{author}{Fujii, N.} \&
  \bibinfo{author}{Graybiel, A.~M.}
\newblock \bibinfo{title}{Neural representation of time in cortico-basal
  ganglia circuits}.
\newblock \emph{\bibinfo{journal}{Proc. Natl. Acad. Sci. USA}}
  \textbf{\bibinfo{volume}{106}}, \bibinfo{pages}{19156--19161}
  (\bibinfo{year}{2009}).

\bibitem{Mita_2009}
\bibinfo{author}{Mita, A.}, \bibinfo{author}{Mushiake, H.},
  \bibinfo{author}{Shima, K.}, \bibinfo{author}{Matsuzaka, Y.} \&
  \bibinfo{author}{Tanji, J.}
\newblock \bibinfo{title}{Interval time coding by neurons in the
  presupplementary and supplementary motor areas}.
\newblock \emph{\bibinfo{journal}{Nat. Neurosci.}}
  \textbf{\bibinfo{volume}{12}}, \bibinfo{pages}{502--507}
  (\bibinfo{year}{2009}).

\bibitem{Wang_2018}
\bibinfo{author}{Wang, J.}, \bibinfo{author}{Narain, D.},
  \bibinfo{author}{Hosseini, E.~A.} \& \bibinfo{author}{Jazayeri, M.}
\newblock \bibinfo{title}{Flexible timing by temporal scaling of cortical
  responses}.
\newblock \emph{\bibinfo{journal}{Nat. Neurosci.}}
  \textbf{\bibinfo{volume}{21}}, \bibinfo{pages}{102--110}
  (\bibinfo{year}{2018}).

\bibitem{Buhusi_2005}
\bibinfo{author}{Buhusi, C.~V.} \& \bibinfo{author}{Meck, W.~H.}
\newblock \bibinfo{title}{What makes us tick? functional and neural mechanisms
  of interval timing}.
\newblock \emph{\bibinfo{journal}{Nat. Rev. Neurosci.}}
  \textbf{\bibinfo{volume}{6}}, \bibinfo{pages}{755--765}
  (\bibinfo{year}{2005}).

\bibitem{Cook_2022}
\bibinfo{author}{Cook, J.~R.} \emph{et~al.}
\newblock \bibinfo{title}{Secondary auditory cortex mediates a sensorimotor
  mechanism for action timing}.
\newblock \emph{\bibinfo{journal}{Nat. Neurosci.}}
  \textbf{\bibinfo{volume}{25}}, \bibinfo{pages}{330--344}
  (\bibinfo{year}{2022}).

\bibitem{Zeki_2019}
\bibinfo{author}{Zeki, M.} \& \bibinfo{author}{Balci, F.}
\newblock \bibinfo{title}{A simplified model of communication between time
  cells: Accounting for the linearly increasing timing imprecision}.
\newblock \emph{\bibinfo{journal}{Front. Comput. Neurosci.}}
  \textbf{\bibinfo{volume}{12}}, \bibinfo{pages}{111} (\bibinfo{year}{2019}).

\bibitem{Matell_2004}
\bibinfo{author}{Matell, M.~S.} \& \bibinfo{author}{Meck, W.~H.}
\newblock \bibinfo{title}{Cortico-striatal circuits and interval timing:
  coincidence detection of oscillatory processes}.
\newblock \emph{\bibinfo{journal}{Cogn. Brain Res.}}
  \textbf{\bibinfo{volume}{21}}, \bibinfo{pages}{139--170}
  (\bibinfo{year}{2004}).

\bibitem{Shuler_2006}
\bibinfo{author}{Shuler, M.~G.} \& \bibinfo{author}{Bear, M.~F.}
\newblock \bibinfo{title}{Reward timing in the primary visual cortex}.
\newblock \emph{\bibinfo{journal}{Science}} \textbf{\bibinfo{volume}{311}},
  \bibinfo{pages}{1606--1609} (\bibinfo{year}{2006}).

\bibitem{Paton_2018}
\bibinfo{author}{Paton, J.~J.} \& \bibinfo{author}{Buonomano, D.~V.}
\newblock \bibinfo{title}{The neural basis of timing: Distributed mechanisms
  for diverse functions}.
\newblock \emph{\bibinfo{journal}{Neuron}} \textbf{\bibinfo{volume}{98}},
  \bibinfo{pages}{687--705} (\bibinfo{year}{2018}).

\bibitem{Ivry_2008}
\bibinfo{author}{Ivry, R.~B.} \& \bibinfo{author}{Schlerf, J.~E.}
\newblock \bibinfo{title}{Dedicated and intrinsic models of time perception}.
\newblock \emph{\bibinfo{journal}{Trends Cogn. Sci.}}
  \textbf{\bibinfo{volume}{12}}, \bibinfo{pages}{1606--1609}
  (\bibinfo{year}{2008}).

\bibitem{Allman_2014}
\bibinfo{author}{Allman, M.~J.}, \bibinfo{author}{Teki, S.},
  \bibinfo{author}{Griffiths, T.~D.} \& \bibinfo{author}{Meck, W.~H.}
\newblock \bibinfo{title}{Properties of the internal clock: First- and
  second-order principles of subjective time}.
\newblock \emph{\bibinfo{journal}{Annu. Rev. Psychol.}}
  \textbf{\bibinfo{volume}{65}}, \bibinfo{pages}{743--771}
  (\bibinfo{year}{2014}).

\bibitem{Hetling_2008}
\bibinfo{author}{Hetling, J.~R.}
\newblock \bibinfo{title}{Comment on 'what is neural engineering?'}.
\newblock \emph{\bibinfo{journal}{J. Neural Eng.}}
  \textbf{\bibinfo{volume}{5}}, \bibinfo{pages}{360} (\bibinfo{year}{2008}).

\bibitem{Poewe_2017}
\bibinfo{author}{Poewe, W.} \emph{et~al.}
\newblock \bibinfo{title}{Parkinson disease}.
\newblock \emph{\bibinfo{journal}{Nat. Rev. Dis. Primers}}
  \textbf{\bibinfo{volume}{3}}, \bibinfo{pages}{17013} (\bibinfo{year}{2017}).

\bibitem{Bertram_2009}
\bibinfo{author}{Bertram, E.~H.}
\newblock \bibinfo{title}{Temporal lobe epilepsy: Where do the seizures really
  begin?}
\newblock \emph{\bibinfo{journal}{Epilepsy Behav.}} \textbf{\bibinfo{volume}{14
  (Suppl 1)}}, \bibinfo{pages}{32--37} (\bibinfo{year}{2009}).

\bibitem{Wu_2019}
\bibinfo{author}{Wu, X.} \emph{et~al.}
\newblock \bibinfo{title}{Altered intrinsic brain activity associated with
  outcome in frontal lobe epilepsy}.
\newblock \emph{\bibinfo{journal}{Sci. Rep.}} \textbf{\bibinfo{volume}{9}},
  \bibinfo{pages}{8989} (\bibinfo{year}{2019}).

\bibitem{Gu_2016}
\bibinfo{author}{Gu, B.-M.}, \bibinfo{author}{Jurkowski, A.~J.},
  \bibinfo{author}{Shi, Z.} \& \bibinfo{author}{Meck, W.~H.}
\newblock \bibinfo{title}{Bayesian optimization of interval timing and biases
  in temporal memory as a function of temporal context, feedback, and dopamine
  levels in young, aged and {Parkinson's} disease patients}.
\newblock \emph{\bibinfo{journal}{Timing Time Percept.}}
  \textbf{\bibinfo{volume}{4}}, \bibinfo{pages}{315--342}
  (\bibinfo{year}{2016}).

\bibitem{Greyson_2014}
\bibinfo{author}{Greyson, B.}, \bibinfo{author}{Fountain, N.~B.},
  \bibinfo{author}{Derr, L.~L.} \& \bibinfo{author}{Broshek, D.~K.}
\newblock \bibinfo{title}{Out-of-body experiences associated with seizures}.
\newblock \emph{\bibinfo{journal}{Front. Hum. Neurosci.}}
  \textbf{\bibinfo{volume}{8}}, \bibinfo{pages}{65} (\bibinfo{year}{2014}).

\bibitem{Cainelli_2019}
\bibinfo{author}{Cainelli, E.}, \bibinfo{author}{Mioni, G.},
  \bibinfo{author}{Boniver, C.}, \bibinfo{author}{Bisiacchi, P.~S.} \&
  \bibinfo{author}{Vecchi, M.}
\newblock \bibinfo{title}{Time perception in childhood absence epilepsy:
  Findings from a pilot study}.
\newblock \emph{\bibinfo{journal}{Epilepsy Behav.}}
  \textbf{\bibinfo{volume}{99}}, \bibinfo{pages}{106460}
  (\bibinfo{year}{2019}).

\bibitem{Patel_2003}
\bibinfo{author}{Patel, A.~D.}
\newblock \bibinfo{title}{Language, music, syntax and the brain}.
\newblock \emph{\bibinfo{journal}{Nat. Neurosci.}}
  \textbf{\bibinfo{volume}{6}}, \bibinfo{pages}{674--681}
  (\bibinfo{year}{2003}).

\bibitem{Janata_2003}
\bibinfo{author}{Janata, P.} \& \bibinfo{author}{Grafton, S.~T.}
\newblock \bibinfo{title}{Swinging in the brain: shared neural substrates for
  behaviors related to sequencing and music}.
\newblock \emph{\bibinfo{journal}{Nat. Neurosci.}}
  \textbf{\bibinfo{volume}{6}}, \bibinfo{pages}{682--687}
  (\bibinfo{year}{2003}).

\bibitem{Hickok_2012}
\bibinfo{author}{Hickok, G.}
\newblock \bibinfo{title}{Computational neuroanatomy of speech production}.
\newblock \emph{\bibinfo{journal}{Nat. Rev. Neurosci.}}
  \textbf{\bibinfo{volume}{13}}, \bibinfo{pages}{135--145}
  (\bibinfo{year}{2012}).

\bibitem{Pool_1954}
\bibinfo{author}{Pool, J.~L.}
\newblock \bibinfo{title}{Psychosurgery in older people}.
\newblock \emph{\bibinfo{journal}{J. Am. Geriatr. Soc.}}
  \textbf{\bibinfo{volume}{2}}, \bibinfo{pages}{456--466}
  (\bibinfo{year}{1954}).

\bibitem{Benabid_1994}
\bibinfo{author}{Benabid, A.~L.} \emph{et~al.}
\newblock \bibinfo{title}{Acute and long-term effects of subthalamic nucleus
  stimulation in {Parkinson's} disease}.
\newblock \emph{\bibinfo{journal}{Stereotact. Funct. Neurosurg.}}
  \textbf{\bibinfo{volume}{62}}, \bibinfo{pages}{76--84}
  (\bibinfo{year}{1994}).

\bibitem{Limousin_1995}
\bibinfo{author}{Limousin, P.} \emph{et~al.}
\newblock \bibinfo{title}{Effect on parkinsonian signs and symptoms of
  bilateral subthalamic nucleus stimulation}.
\newblock \emph{\bibinfo{journal}{Lancet}} \textbf{\bibinfo{volume}{62}},
  \bibinfo{pages}{91--95} (\bibinfo{year}{1995}).

\bibitem{Siegfried_1994}
\bibinfo{author}{Siegfried, J.} \& \bibinfo{author}{Lippitz, B.}
\newblock \bibinfo{title}{Bilateral chronic electrostimulation of
  ventroposterolateral pallidum: A new therapeutic approach for alleviating all
  parkinsonian symptoms}.
\newblock \emph{\bibinfo{journal}{Neurosurgery}} \textbf{\bibinfo{volume}{35}},
  \bibinfo{pages}{1126--1129} (\bibinfo{year}{1994}).

\bibitem{Bouthour_2019}
\bibinfo{author}{Bouthour, W.} \emph{et~al.}
\newblock \bibinfo{title}{Biomarkers for closed-loop deep brain stimulation in
  {Parkinson} disease and beyond}.
\newblock \emph{\bibinfo{journal}{Nat. Rev. Neurol.}}
  \textbf{\bibinfo{volume}{15}}, \bibinfo{pages}{343--352}
  (\bibinfo{year}{2019}).

\bibitem{Krauss_2021}
\bibinfo{author}{Krauss, J.~K.} \emph{et~al.}
\newblock \bibinfo{title}{Technology of deep brain stimulation: current status
  and future directions}.
\newblock \emph{\bibinfo{journal}{Nat. Rev. Neurol.}}
  \textbf{\bibinfo{volume}{17}}, \bibinfo{pages}{75--87}
  (\bibinfo{year}{2021}).

\bibitem{Okun_2012}
\bibinfo{author}{Okun, M.~S.}
\newblock \bibinfo{title}{Deep-brain stimulation for {Parkinson's} disease}.
\newblock \emph{\bibinfo{journal}{N. Engl. J. Med.}}
  \textbf{\bibinfo{volume}{367}}, \bibinfo{pages}{1529--1538}
  (\bibinfo{year}{2012}).

\bibitem{Rizzone_2001}
\bibinfo{author}{Rizzone, M.} \emph{et~al.}
\newblock \bibinfo{title}{Deep brain stimulation of the subthalamic nucleus in
  {Parkinson's} disease: effects of variation in stimulation parameters}.
\newblock \emph{\bibinfo{journal}{J. Neurol. Neurosurg. Psychiatry}}
  \textbf{\bibinfo{volume}{71}}, \bibinfo{pages}{215--219}
  (\bibinfo{year}{2001}).

\bibitem{Kuncel_2006}
\bibinfo{author}{Kuncel, A.~M.} \emph{et~al.}
\newblock \bibinfo{title}{Clinical response to varying the stimulus parameters
  in deep brain stimulation for essential tremor}.
\newblock \emph{\bibinfo{journal}{Mov. Disord.}} \textbf{\bibinfo{volume}{21}},
  \bibinfo{pages}{1920--1928} (\bibinfo{year}{2006}).

\bibitem{Oswal_2013}
\bibinfo{author}{Oswal, A.}, \bibinfo{author}{Brown, P.} \&
  \bibinfo{author}{Litvak, V.}
\newblock \bibinfo{title}{Synchronized neural oscillations and the
  pathophysiology of {Parkinson's} disease}.
\newblock \emph{\bibinfo{journal}{Curr. Opin. Neurol.}}
  \textbf{\bibinfo{volume}{26}}, \bibinfo{pages}{662--670}
  (\bibinfo{year}{2013}).

\bibitem{Cheyne_2013}
\bibinfo{author}{Cheyne, D.~O.}
\newblock \bibinfo{title}{{MEG} studies of sensorimotor rhythms: a review}.
\newblock \emph{\bibinfo{journal}{Exp. Neurol.}}
  \textbf{\bibinfo{volume}{245}}, \bibinfo{pages}{27--39}
  (\bibinfo{year}{2013}).

\bibitem{Shah_2018}
\bibinfo{author}{Shah, S.~A.}, \bibinfo{author}{Tinkhauser, G.},
  \bibinfo{author}{Chen, C.~C.}, \bibinfo{author}{Little, S.} \&
  \bibinfo{author}{Brown, P.}
\newblock \bibinfo{title}{Parkinsonian tremor detection from subthalamic
  nucleus local field potentials for closed-loop deep brain stimulation}.
\newblock In \emph{\bibinfo{booktitle}{Conf. Proc. IEEE Eng. Med. Biol. Soc.}},
  \bibinfo{pages}{2320--2324} (\bibinfo{year}{2018}).

\bibitem{Tan_2019}
\bibinfo{author}{Tan, H.} \emph{et~al.}
\newblock \bibinfo{title}{Decoding voluntary movements and postural tremor
  based on thalamic {LFPs} as a basis for closed-loop stimulation for essential
  tremor}.
\newblock \emph{\bibinfo{journal}{Brain Stimul.}}
  \textbf{\bibinfo{volume}{12}}, \bibinfo{pages}{858--867}
  (\bibinfo{year}{2019}).

\bibitem{Scangos_2021b}
\bibinfo{author}{Scangos, K.~W.} \emph{et~al.}
\newblock \bibinfo{title}{Closed-loop neuromodulation in an individual with
  treatment-resistant depression}.
\newblock \emph{\bibinfo{journal}{Nat. Med.}} \textbf{\bibinfo{volume}{27}},
  \bibinfo{pages}{1696--1700} (\bibinfo{year}{2021}).

\bibitem{Scangos_2021a}
\bibinfo{author}{Scangos, K.~W.}, \bibinfo{author}{Makhoul, G.~S.},
  \bibinfo{author}{Sugrue, L.~P.}, \bibinfo{author}{Chang, E.~F.} \&
  \bibinfo{author}{Krystal, A.~D.}
\newblock \bibinfo{title}{State-dependent responses to intracranial brain
  stimulation in a patient with depression}.
\newblock \emph{\bibinfo{journal}{Nat. Med.}} \textbf{\bibinfo{volume}{27}},
  \bibinfo{pages}{229--231} (\bibinfo{year}{2021}).

\bibitem{Cao_2022}
\bibinfo{author}{Cao, M.} \emph{et~al.}
\newblock \bibinfo{title}{Virtual intracranial {EEG} signals reconstructed from
  {MEG} with potential for epilepsy surgery}.
\newblock \emph{\bibinfo{journal}{Nat. Commun.}} \textbf{\bibinfo{volume}{13}},
  \bibinfo{pages}{994} (\bibinfo{year}{2022}).

\bibitem{Sinha_2017}
\bibinfo{author}{Sinha, N.} \emph{et~al.}
\newblock \bibinfo{title}{Predicting neurosurgical outcomes in focal epilepsy
  patients using computational modelling}.
\newblock \emph{\bibinfo{journal}{Brain}} \textbf{\bibinfo{volume}{140}},
  \bibinfo{pages}{319--332} (\bibinfo{year}{2017}).

\bibitem{Pasley_2012}
\bibinfo{author}{Pasley, B.~N.} \emph{et~al.}
\newblock \bibinfo{title}{Reconstructing speech from human auditory cortex}.
\newblock \emph{\bibinfo{journal}{PLoS Biol.}} \textbf{\bibinfo{volume}{10}},
  \bibinfo{pages}{e1001251} (\bibinfo{year}{2012}).

\bibitem{Angrick_2019}
\bibinfo{author}{Angrick, M.} \emph{et~al.}
\newblock \bibinfo{title}{Speech synthesis from {ECoG} using densely connected
  {3D} convolutional neural networks}.
\newblock \emph{\bibinfo{journal}{J. Neural Eng.}}
  \textbf{\bibinfo{volume}{16}}, \bibinfo{pages}{036019}
  (\bibinfo{year}{2019}).

\bibitem{Makin_2020}
\bibinfo{author}{Makin, J.~G.}, \bibinfo{author}{Moses, D.~A.} \&
  \bibinfo{author}{Chang, E.~F.}
\newblock \bibinfo{title}{Machine translation of cortical activity to text with
  an encoder-decoder framework}.
\newblock \emph{\bibinfo{journal}{Nat. Neurosci.}}
  \textbf{\bibinfo{volume}{23}}, \bibinfo{pages}{575--582}
  (\bibinfo{year}{2020}).

\bibitem{Cogan_2020}
\bibinfo{author}{Cogan, G.~B.}
\newblock \bibinfo{title}{Translating the brain}.
\newblock \emph{\bibinfo{journal}{Nat. Neurosci.}}
  \textbf{\bibinfo{volume}{23}}, \bibinfo{pages}{469--472}
  (\bibinfo{year}{2020}).

\bibitem{Moses_2019}
\bibinfo{author}{Moses, D.~A.}, \bibinfo{author}{Leonard, M.~K.},
  \bibinfo{author}{Makin, J.~G.} \& \bibinfo{author}{Chang, E.~F.}
\newblock \bibinfo{title}{Real-time decoding of question-and-answer speech
  dialogue using human cortical activity}.
\newblock \emph{\bibinfo{journal}{Nat. Commun.}} \textbf{\bibinfo{volume}{10}},
  \bibinfo{pages}{3096} (\bibinfo{year}{2019}).

\bibitem{Hassabis_2017}
\bibinfo{author}{Hassabis, D.}, \bibinfo{author}{Kumaran, D.},
  \bibinfo{author}{Summerfield, C.} \& \bibinfo{author}{Botvinick, M.}
\newblock \bibinfo{title}{Neuroscience-inspired artificial intelligence}.
\newblock \emph{\bibinfo{journal}{Neuron}} \textbf{\bibinfo{volume}{95}},
  \bibinfo{pages}{245--258} (\bibinfo{year}{2017}).

\bibitem{Frenkel_2021}
\bibinfo{author}{Frenkel, C.}
\newblock \bibinfo{title}{Sparsity provides a competitive advantage}.
\newblock \emph{\bibinfo{journal}{Nat. Mach. Intell.}}
  \textbf{\bibinfo{volume}{3}}, \bibinfo{pages}{742--743}
  (\bibinfo{year}{2021}).

\bibitem{Dayan_2001}
\bibinfo{author}{Dayan, P.} \& \bibinfo{author}{Abbott, L.~F.}
\newblock \emph{\bibinfo{title}{Theoretical neuroscience: computational and
  mathematical modeling of neural systems}} (\bibinfo{publisher}{The MIT
  Press}, \bibinfo{address}{Cambridge}, \bibinfo{year}{2001}).

\bibitem{Beniaguev_2021}
\bibinfo{author}{Beniaguev, D.}, \bibinfo{author}{Segev, I.} \&
  \bibinfo{author}{London, M.}
\newblock \bibinfo{title}{Single cortical neurons as deep artificial neural
  networks}.
\newblock \emph{\bibinfo{journal}{Neuron}} \textbf{\bibinfo{volume}{109}},
  \bibinfo{pages}{2727--2739} (\bibinfo{year}{2021}).

\bibitem{Lechner_2020}
\bibinfo{author}{Lechner, M.} \emph{et~al.}
\newblock \bibinfo{title}{Neural circuit policies enabling auditable autonomy}.
\newblock \emph{\bibinfo{journal}{Nat. Mach. Intell.}}
  \textbf{\bibinfo{volume}{2}}, \bibinfo{pages}{642--652}
  (\bibinfo{year}{2020}).

\bibitem{OConnor_2005}
\bibinfo{author}{O'Connor, D.~H.}, \bibinfo{author}{Wittenberg, G.~M.} \&
  \bibinfo{author}{Wang, S. S.-H.}
\newblock \bibinfo{title}{Graded bidirectional synaptic plasticity is composed
  of switch-like unitary events}.
\newblock \emph{\bibinfo{journal}{Proc. Natl. Acad. Sci. USA}}
  \textbf{\bibinfo{volume}{102}}, \bibinfo{pages}{9679--9684}
  (\bibinfo{year}{2005}).

\bibitem{Courbariaux_2016}
\bibinfo{author}{Courbariaux, M.}, \bibinfo{author}{Hubara, I.},
  \bibinfo{author}{Soudry, D.}, \bibinfo{author}{El-Yaniv, R.} \&
  \bibinfo{author}{Bengio, Y.}
\newblock \bibinfo{title}{Binarized neural networks: Training deep neural
  networks with weights and activations constrained to +1 or -1}.
\newblock \emph{\bibinfo{journal}{arXiv:1602.02830}}  (\bibinfo{year}{2016}).

\bibitem{Graupner_2010}
\bibinfo{author}{Graupner, M.} \& \bibinfo{author}{Brunel, N.}
\newblock \bibinfo{title}{Mechanisms of induction and maintenance of
  spike-timing dependent plasticity in biophysical synapse models}.
\newblock \emph{\bibinfo{journal}{Front. Comput. Neurosci.}}
  \textbf{\bibinfo{volume}{4}}, \bibinfo{pages}{136} (\bibinfo{year}{2010}).

\bibitem{Baldassi_2007}
\bibinfo{author}{Baldassi, C.}, \bibinfo{author}{Braunstein, A.},
  \bibinfo{author}{Brunel, N.} \& \bibinfo{author}{Zecchina, R.}
\newblock \bibinfo{title}{Efficient supervised learning in networks with binary
  synapses}.
\newblock \emph{\bibinfo{journal}{Proc. Natl. Acad. Sci. USA}}
  \textbf{\bibinfo{volume}{104}}, \bibinfo{pages}{11079--11084}
  (\bibinfo{year}{2007}).

\bibitem{Kirkpatrick_2017}
\bibinfo{author}{Kirkpatrick, J.} \emph{et~al.}
\newblock \bibinfo{title}{Overcoming catastrophic forgetting in neural
  networks}.
\newblock \emph{\bibinfo{journal}{Proc. Natl. Acad. Sci. USA}}
  \textbf{\bibinfo{volume}{114}}, \bibinfo{pages}{3521--3526}
  (\bibinfo{year}{2017}).

\bibitem{Ji_2007}
\bibinfo{author}{Ji, D.} \& \bibinfo{author}{Wilson, M.}
\newblock \bibinfo{title}{Coordinated memory replay in the visual cortex and
  hippocampus during sleep}.
\newblock \emph{\bibinfo{journal}{Nat. Neurosci.}}
  \textbf{\bibinfo{volume}{10}}, \bibinfo{pages}{100--107}
  (\bibinfo{year}{2007}).

\bibitem{Mnih_2015}
\bibinfo{author}{Mnih, V.} \emph{et~al.}
\newblock \bibinfo{title}{Human-level control through deep reinforcement
  learning}.
\newblock \emph{\bibinfo{journal}{Nature}} \textbf{\bibinfo{volume}{518}},
  \bibinfo{pages}{529--533} (\bibinfo{year}{2015}).

\bibitem{Sutton_2018}
\bibinfo{author}{Sutton, R.~S.} \& \bibinfo{author}{Barto, A.~G.}
\newblock \emph{\bibinfo{title}{Reinforcement Learning: An Introduction}}
  (\bibinfo{publisher}{The MIT Press}, \bibinfo{address}{Cambridge},
  \bibinfo{year}{2018}).

\bibitem{Salinas_2000}
\bibinfo{author}{Salinas, E.} \& \bibinfo{author}{Thier, P.}
\newblock \bibinfo{title}{Gain modulation: A major computational principle of
  the central nervous system}.
\newblock \emph{\bibinfo{journal}{Neuron}} \textbf{\bibinfo{volume}{27}},
  \bibinfo{pages}{15--21} (\bibinfo{year}{2000}).

\bibitem{Salinas_2001}
\bibinfo{author}{Salinas, E.} \& \bibinfo{author}{Sejnowski, T.~J.}
\newblock \bibinfo{title}{Gain modulation in the central nervous system: Where
  behavior, neurophysiology, and computation meet}.
\newblock \emph{\bibinfo{journal}{Neuroscientist}}
  \textbf{\bibinfo{volume}{7}}, \bibinfo{pages}{430--440}
  (\bibinfo{year}{2001}).

\bibitem{Vaswani_2017}
\bibinfo{author}{Vaswani, A.} \emph{et~al.}
\newblock \bibinfo{title}{Attention is all you need}.
\newblock \emph{\bibinfo{journal}{arXiv:1706.03762}}  (\bibinfo{year}{2017}).

\bibitem{Devlin_2019}
\bibinfo{author}{Devlin, J.}, \bibinfo{author}{Chang, M.-W.},
  \bibinfo{author}{Lee, K.} \& \bibinfo{author}{Toutanova, K.}
\newblock \bibinfo{title}{{BERT}: Pre-training of deep bidirectional
  transformers for language understanding}.
\newblock \emph{\bibinfo{journal}{arXiv:1810.04805}}  (\bibinfo{year}{2019}).

\bibitem{Cichon_2015}
\bibinfo{author}{Cichon, J.} \& \bibinfo{author}{Gan, W.-B.}
\newblock \bibinfo{title}{Branch-specific dendritic ca2+ spikes cause
  persistent synaptic plasticity}.
\newblock \emph{\bibinfo{journal}{Nature}} \textbf{\bibinfo{volume}{520}},
  \bibinfo{pages}{180--185} (\bibinfo{year}{2015}).

\bibitem{Manning_2020}
\bibinfo{author}{Manning, C.~D.}, \bibinfo{author}{Clark, K.},
  \bibinfo{author}{Hewitt, J.}, \bibinfo{author}{Khandelwal, U.} \&
  \bibinfo{author}{Levy, O.}
\newblock \bibinfo{title}{Emergent linguistic structure in artificial neural
  networks trained by self-supervision}.
\newblock \emph{\bibinfo{journal}{Proc. Natl. Acad. Sci. USA}}
  \textbf{\bibinfo{volume}{117}}, \bibinfo{pages}{30046--30054}
  (\bibinfo{year}{2020}).

\bibitem{Zeng_2019}
\bibinfo{author}{Zeng, G.}, \bibinfo{author}{Chen, Y.}, \bibinfo{author}{Cui,
  B.} \& \bibinfo{author}{Yu, S.}
\newblock \bibinfo{title}{Continual learning of context-dependent processing in
  neural networks}.
\newblock \emph{\bibinfo{journal}{Nat. Mach. Intell.}}
  \textbf{\bibinfo{volume}{1}}, \bibinfo{pages}{369--372}
  (\bibinfo{year}{2019}).

\bibitem{Lillicrap_2020}
\bibinfo{author}{Lillicrap, T.~P.}, \bibinfo{author}{Santoro, A.},
  \bibinfo{author}{Marris, L.}, \bibinfo{author}{Akerman, C.~J.} \&
  \bibinfo{author}{Hinton, G.}
\newblock \bibinfo{title}{Backpropagation and the brain}.
\newblock \emph{\bibinfo{journal}{Nat. Rev. Neurosci.}}
  \textbf{\bibinfo{volume}{21}}, \bibinfo{pages}{335--346}
  (\bibinfo{year}{2020}).

\bibitem{Hong_2016}
\bibinfo{author}{Hong, H.}, \bibinfo{author}{Yamins, D. L.~K.},
  \bibinfo{author}{Majaj, N.~J.} \& \bibinfo{author}{DiCarlo, J.~J.}
\newblock \bibinfo{title}{Explicit information for categoryorthogonal object
  properties increases along the ventral stream}.
\newblock \emph{\bibinfo{journal}{Nat. Neurosci.}}
  \textbf{\bibinfo{volume}{19}}, \bibinfo{pages}{613--622}
  (\bibinfo{year}{2016}).

\bibitem{Goldstein_2022}
\bibinfo{author}{Goldstein, A.} \emph{et~al.}
\newblock \bibinfo{title}{Shared computational principles for language
  processing in humans and deep language models}.
\newblock \emph{\bibinfo{journal}{Nat. Neurosci.}}
  \textbf{\bibinfo{volume}{25}}, \bibinfo{pages}{369--380}
  (\bibinfo{year}{2022}).

\bibitem{Mante_2013}
\bibinfo{author}{Mante, V.}, \bibinfo{author}{Sussillo, D.},
  \bibinfo{author}{Shenoy, K.~V.} \& \bibinfo{author}{Newsome, W.~T.}
\newblock \bibinfo{title}{Context-dependent computation by recurrent dynamics
  in prefrontal cortex}.
\newblock \emph{\bibinfo{journal}{Nature}} \textbf{\bibinfo{volume}{503}},
  \bibinfo{pages}{78--84} (\bibinfo{year}{2013}).

\bibitem{Bi_2022}
\bibinfo{author}{Bi, Z.}, \bibinfo{author}{Chen, G.}, \bibinfo{author}{Yang,
  D.}, \bibinfo{author}{Zhou, Y.} \& \bibinfo{author}{Tian, L.}
\newblock \bibinfo{title}{Evolutionary learning in the brain by heterosynaptic
  plasticity}.
\newblock \emph{\bibinfo{journal}{bioRxiv:2021.12.14.472260}}
  (\bibinfo{year}{2022}).

\bibitem{Baldassi_2015}
\bibinfo{author}{Baldassi, C.}, \bibinfo{author}{Ingrosso, A.},
  \bibinfo{author}{Lucibello, C.}, \bibinfo{author}{Saglietti, L.} \&
  \bibinfo{author}{Zecchina, R.}
\newblock \bibinfo{title}{Subdominant dense clusters allow for simple learning
  and high computational performance in neural networks with discrete
  synapses}.
\newblock \emph{\bibinfo{journal}{Phys. Rev. Lett.}}
  \textbf{\bibinfo{volume}{115}}, \bibinfo{pages}{128101}
  (\bibinfo{year}{2015}).

\bibitem{Bi_2020b}
\bibinfo{author}{Bi, Z.} \& \bibinfo{author}{Zhou, C.}
\newblock \bibinfo{title}{Understanding the computational difficulty of a
  binary-weight perceptron and the advantage of input sparseness}.
\newblock \emph{\bibinfo{journal}{J. Phys. A: Math. Theor.}}
  \textbf{\bibinfo{volume}{53}}, \bibinfo{pages}{035002}
  (\bibinfo{year}{2020}).

\bibitem{Herculano-Houzel_2012}
\bibinfo{author}{Herculano-Houzel, S.}
\newblock \bibinfo{title}{The remarkable, yet not extraordinary, human brain as
  a scaled-up primate brain and its associated cost}.
\newblock \emph{\bibinfo{journal}{Proc. Natl. Acad. Sci. USA}}
  \textbf{\bibinfo{volume}{109}}, \bibinfo{pages}{10661--10668}
  (\bibinfo{year}{2012}).

\bibitem{Brown_2020}
\bibinfo{author}{Brown, T.~B.} \emph{et~al.}
\newblock \bibinfo{title}{Language models are few-shot learners}.
\newblock \emph{\bibinfo{journal}{arXiv:2005.14165}}  (\bibinfo{year}{2020}).

\bibitem{Bommasani_2021}
\bibinfo{author}{Bommasani, R.} \emph{et~al.}
\newblock \bibinfo{title}{On the opportunities and risks of foundation models}.
\newblock \emph{\bibinfo{journal}{arXiv:2108.07258}}  (\bibinfo{year}{2021}).

\bibitem{Rives_2021}
\bibinfo{author}{Rives, A.} \emph{et~al.}
\newblock \bibinfo{title}{Biological structure and function emerge from scaling
  unsupervised learning to 250 million protein sequences}.
\newblock \emph{\bibinfo{journal}{Proc. Natl. Acad. Sci. USA}}
  \textbf{\bibinfo{volume}{118}}, \bibinfo{pages}{e2016239118}
  (\bibinfo{year}{2021}).

\bibitem{Ramesh_2021}
\bibinfo{author}{Ramesh, A.} \emph{et~al.}
\newblock \bibinfo{title}{Zero-shot text-to-image generation}.
\newblock \emph{\bibinfo{journal}{arXiv:2102.12092}}  (\bibinfo{year}{2021}).

\bibitem{Schwaller_2021}
\bibinfo{author}{Schwaller, P.} \emph{et~al.}
\newblock \bibinfo{title}{Mapping the space of chemical reactions using
  attention-based neural networks}.
\newblock \emph{\bibinfo{journal}{Nat. Mach. Intell.}}
  \textbf{\bibinfo{volume}{3}}, \bibinfo{pages}{144--152}
  (\bibinfo{year}{2021}).

\bibitem{Strubell_2019}
\bibinfo{author}{Strubell, E.}, \bibinfo{author}{Ganesh, A.} \&
  \bibinfo{author}{McCallum, A.}
\newblock \bibinfo{title}{Energy and policy considerations for deep learning in
  nlp}.
\newblock \emph{\bibinfo{journal}{arXiv:1906.02243}}  (\bibinfo{year}{2019}).

\bibitem{Nielsen_2011}
\bibinfo{author}{Nielsen, M.~A.} \& \bibinfo{author}{Chuang, I.~L.}
\newblock \emph{\bibinfo{title}{Quantum Computation and Quantum Information}}
  (\bibinfo{publisher}{Cambridge University Press},
  \bibinfo{address}{Cambridge}, \bibinfo{year}{2011}).

\bibitem{Morita_2020}
\bibinfo{author}{Morita, K.} \emph{et~al.}
\newblock \bibinfo{title}{Clonal evolution of acute myeloid leukemia revealed
  by high-throughput single-cell genomics}.
\newblock \emph{\bibinfo{journal}{Nat. Commun.}} \textbf{\bibinfo{volume}{11}},
  \bibinfo{pages}{5327} (\bibinfo{year}{2020}).

\bibitem{Black_2021}
\bibinfo{author}{Black, J. R.~M.} \& \bibinfo{author}{McGranahan, N.}
\newblock \bibinfo{title}{Genetic and non-genetic clonal diversity in cancer
  evolution}.
\newblock \emph{\bibinfo{journal}{Nat. Rev. Cancer}}
  \textbf{\bibinfo{volume}{21}}, \bibinfo{pages}{379--392}
  (\bibinfo{year}{2021}).

\bibitem{Swanton_2011}
\bibinfo{author}{Swanton, D.~N.} \& \bibinfo{author}{Matell, M.~S.}
\newblock \bibinfo{title}{Stimulus compounding in interval timing: The
  modality-duration relationship of the anchor durations results in
  qualitatively different response patterns to the compound cue}.
\newblock \emph{\bibinfo{journal}{J. Exp. Psychol. Anim. Behav. Process}}
  \textbf{\bibinfo{volume}{37}}, \bibinfo{pages}{94--107}
  (\bibinfo{year}{2011}).

\bibitem{Matell_2014}
\bibinfo{author}{Matell, M.~S.} \& \bibinfo{author}{Kurti, A.~N.}
\newblock \bibinfo{title}{Reinforcement probability modulates temporal memory
  selection and integration processes}.
\newblock \emph{\bibinfo{journal}{Acta Psychol. (Amst)}}
  \textbf{\bibinfo{volume}{147}}, \bibinfo{pages}{80--91}
  (\bibinfo{year}{2014}).

\bibitem{Matell_2017}
\bibinfo{author}{Matell, M.~S.} \& \bibinfo{author}{Valle, R. B.~D.}
\newblock \bibinfo{title}{Temporal specificity in pavlovian-to-instrumental
  transfer}.
\newblock \emph{\bibinfo{journal}{Learn Mem.}} \textbf{\bibinfo{volume}{25}},
  \bibinfo{pages}{8--20} (\bibinfo{year}{2017}).

\bibitem{Karmarkar_2007}
\bibinfo{author}{Karmarkar, U.~R.} \& \bibinfo{author}{Buonomano, D.~V.}
\newblock \bibinfo{title}{Timing in the absence of clocks: Encoding time in
  neural network states}.
\newblock \emph{\bibinfo{journal}{Neuron}} \textbf{\bibinfo{volume}{53}},
  \bibinfo{pages}{427--438} (\bibinfo{year}{2007}).

\bibitem{Teki_2011}
\bibinfo{author}{Teki, S.}, \bibinfo{author}{Grube, M.},
  \bibinfo{author}{Kumar, S.} \& \bibinfo{author}{Griffiths, T.~D.}
\newblock \bibinfo{title}{Distinct neural substrates of durationbased and
  beat-based auditory timing}.
\newblock \emph{\bibinfo{journal}{J. Neurosci.}} \textbf{\bibinfo{volume}{31}},
  \bibinfo{pages}{3805--3812} (\bibinfo{year}{2011}).

\bibitem{Bruggemann_2016}
\bibinfo{author}{Br{\"u}ggemann, J.} \& \bibinfo{author}{Bizer, K.}
\newblock \bibinfo{title}{Laboratory experiments in innovation research: a
  methodological overview and a review of the current literature}.
\newblock \emph{\bibinfo{journal}{J. Innov. Entrep.}}
  \textbf{\bibinfo{volume}{5}}, \bibinfo{pages}{24} (\bibinfo{year}{2016}).

\bibitem{Hulstijn_1997}
\bibinfo{author}{Hulstijn, J.~H.}
\newblock \bibinfo{title}{Second language acquisition research in the
  laboratory: possibilities and limitations}.
\newblock \emph{\bibinfo{journal}{Stud. Second. Lang. Acquis.}}
  \textbf{\bibinfo{volume}{19}}, \bibinfo{pages}{131--143}
  (\bibinfo{year}{1997}).

\bibitem{Baker_2016}
\bibinfo{author}{Baker, A.}
\newblock \bibinfo{title}{Simplicity}.
\newblock In \bibinfo{editor}{Zalta, E.~N.} (ed.) \emph{\bibinfo{booktitle}{The
  Stanford Encyclopedia of Philosophy}} (\bibinfo{publisher}{Stanford
  University}, \bibinfo{address}{California}, \bibinfo{year}{2016}).

\bibitem{Sober_1991}
\bibinfo{author}{Sober, E.} \& \bibinfo{author}{Knowles, D.}
\newblock \emph{\bibinfo{title}{Let's Razor Ockham's Razor}}.
\newblock Royal Institute of Philosophy Supplements
  (\bibinfo{publisher}{Cambridge University Press}, \bibinfo{year}{1991}).

\bibitem{Popper_1959}
\bibinfo{author}{Popper, K.}
\newblock \emph{\bibinfo{title}{The Logic of Scientific Discovery}}
  (\bibinfo{publisher}{Hutchinson}, \bibinfo{address}{London},
  \bibinfo{year}{1959}).

\bibitem{Wong_2006}
\bibinfo{author}{Wong, K.-F.} \& \bibinfo{author}{Wang, X.-J.}
\newblock \bibinfo{title}{A recurrent network mechanism of time integration in
  perceptual decisions}.
\newblock \emph{\bibinfo{journal}{J. Neurosci.}} \textbf{\bibinfo{volume}{26}},
  \bibinfo{pages}{1314--1328} (\bibinfo{year}{2006}).

\bibitem{Lim_2013}
\bibinfo{author}{Lim, S.} \& \bibinfo{author}{Goldman, M.~S.}
\newblock \bibinfo{title}{Balanced cortical microcircuitry for maintaining
  information in working memory}.
\newblock \emph{\bibinfo{journal}{Nat. Neurosci.}}
  \textbf{\bibinfo{volume}{16}}, \bibinfo{pages}{1306--1314}
  (\bibinfo{year}{2013}).

\bibitem{Hogan_2021}
\bibinfo{author}{Hogan, A.} \emph{et~al.}
\newblock \bibinfo{title}{Knowledge graphs}.
\newblock \emph{\bibinfo{journal}{arXiv:2003.02320}}  (\bibinfo{year}{2021}).

\bibitem{Guo_2022}
\bibinfo{author}{Guo, Q.} \emph{et~al.}
\newblock \bibinfo{title}{A survey on knowledge graph-based recommender
  systems}.
\newblock \emph{\bibinfo{journal}{IEEE Trans. Knowl. Data Eng.}}
  \textbf{\bibinfo{volume}{34}}, \bibinfo{pages}{3549--3568}
  (\bibinfo{year}{2022}).

\bibitem{Annervaz_2018}
\bibinfo{author}{Annervaz, K.~M.}, \bibinfo{author}{Chowdhury, S. B.~R.} \&
  \bibinfo{author}{Dukkipati, A.}
\newblock \bibinfo{title}{Learning beyond datasets: Knowledge graph augmented
  neural networks for natural language processing}.
\newblock \emph{\bibinfo{journal}{arXiv:1802.05930}}  (\bibinfo{year}{2018}).

\bibitem{Zeng_2021}
\bibinfo{author}{Zeng, W.} \emph{et~al.}
\newblock \bibinfo{title}{Pangu-$\alpha$: Large-scale autoregressive pretrained
  chinese language models with auto-parallel computation}.
\newblock \emph{\bibinfo{journal}{arXiv:2104.12369}}  (\bibinfo{year}{2021}).

\bibitem{Colon-Hernandez_2021}
\bibinfo{author}{Colon-Hernandez, P.}, \bibinfo{author}{Havasi, C.},
  \bibinfo{author}{Alonso, J.}, \bibinfo{author}{Huggins, M.} \&
  \bibinfo{author}{Breazeal, C.}
\newblock \bibinfo{title}{Combining pre-trained language models and structured
  knowledge}.
\newblock \emph{\bibinfo{journal}{arXiv:2101.12294}}  (\bibinfo{year}{2021}).

\bibitem{David_Hume}
\bibinfo{title}{{David Hume.}}
  \urlprefix\url{https://en.wikipedia.org/wiki/David_Hume}.

\bibitem{Immanuel_Kant}
\bibinfo{title}{{Immanuel Kant.}}
  \urlprefix\url{https://en.wikipedia.org/wiki/Immanuel_Kant}.

\bibitem{wiki_knowability}
\bibinfo{title}{Fitch's paradox of knowability.}
  \urlprefix\url{https://en.wikipedia.org/wiki/Fitch%27s_paradox_of_knowability}.

\bibitem{Cubitt_2012a}
\bibinfo{author}{Cubitt, T.~S.}, \bibinfo{author}{Eisert, J.} \&
  \bibinfo{author}{Wolf, M.~M.}
\newblock \bibinfo{title}{Extracting dynamical equations from experimental data
  is {NP} hard}.
\newblock \emph{\bibinfo{journal}{Phys. Rev. Lett.}}
  \textbf{\bibinfo{volume}{108}}, \bibinfo{pages}{120503}
  (\bibinfo{year}{2012}).

\bibitem{Cubitt_2012b}
\bibinfo{author}{Cubitt, T.~S.}, \bibinfo{author}{Eisert, J.} \&
  \bibinfo{author}{Wolf, M.~M.}
\newblock \bibinfo{title}{The complexity of relating quantum channels to master
  equations}.
\newblock \emph{\bibinfo{journal}{Commun. Math. Phys.}}
  \textbf{\bibinfo{volume}{310}}, \bibinfo{pages}{383--418}
  (\bibinfo{year}{2012}).

\bibitem{Mezard_2009}
\bibinfo{author}{M{\'e}zard, M.} \& \bibinfo{author}{Montanari, A.}
\newblock \emph{\bibinfo{title}{Information, Physics, and Computation}}
  (\bibinfo{publisher}{Oxford University Press}, \bibinfo{year}{2009}).

\bibitem{Goto_2021}
\bibinfo{author}{Goto, A.} \emph{et~al.}
\newblock \bibinfo{title}{Stepwise synaptic plasticity events drive the early
  phase of memory consolidation}.
\newblock \emph{\bibinfo{journal}{Science}} \textbf{\bibinfo{volume}{374}},
  \bibinfo{pages}{857--863} (\bibinfo{year}{2021}).

\bibitem{Barron_2020}
\bibinfo{author}{Barron, H.~C.} \emph{et~al.}
\newblock \bibinfo{title}{Neuronal computation underlying inferential reasoning
  in humans and mice}.
\newblock \emph{\bibinfo{journal}{Cell}} \textbf{\bibinfo{volume}{183}},
  \bibinfo{pages}{228--243} (\bibinfo{year}{2020}).

\bibitem{Sosa_2021}
\bibinfo{author}{Sosa, M.} \& \bibinfo{author}{Giocomo, L.~M.}
\newblock \bibinfo{title}{Navigating for reward}.
\newblock \emph{\bibinfo{journal}{Nat. Rev. Neurosci.}}
  \textbf{\bibinfo{volume}{22}}, \bibinfo{pages}{472--487}
  (\bibinfo{year}{2021}).

\bibitem{Eichenbaum_2014}
\bibinfo{author}{Eichenbaum, H.}
\newblock \bibinfo{title}{Time cells in the hippocampus: a new dimension for
  mapping memories}.
\newblock \emph{\bibinfo{journal}{Nat. Rev. Neurosci.}}
  \textbf{\bibinfo{volume}{15}}, \bibinfo{pages}{732--744}
  (\bibinfo{year}{2014}).

\bibitem{Turk-Browne_2019}
\bibinfo{author}{Turk-Browne, N.~B.}
\newblock \bibinfo{title}{The hippocampus as a visual area organized by space
  and time: A spatiotemporal similarity hypothesis}.
\newblock \emph{\bibinfo{journal}{Vision Res.}} \textbf{\bibinfo{volume}{165}},
  \bibinfo{pages}{123--130} (\bibinfo{year}{2019}).

\bibitem{Nieh_2021}
\bibinfo{author}{Nieh, E.~H.} \emph{et~al.}
\newblock \bibinfo{title}{Geometry of abstract learned knowledge in the
  hippocampus}.
\newblock \emph{\bibinfo{journal}{Nature}} \textbf{\bibinfo{volume}{595}},
  \bibinfo{pages}{80--84} (\bibinfo{year}{2021}).

\bibitem{wiki_cave}
\bibinfo{title}{Allegory of the cave.}
  \urlprefix\url{https://en.wikipedia.org/wiki/Allegory_of_the_cave}.

\bibitem{Jia_2019}
\bibinfo{author}{Jia, W.~W.}
\newblock \emph{\bibinfo{title}{Dark knowledge: how machine cognition subverts
  business and society}} (\bibinfo{publisher}{CITIC Press Group},
  \bibinfo{address}{Beijing}, \bibinfo{year}{2019}).

\bibitem{Hinton_2015}
\bibinfo{author}{Hinton, G.}, \bibinfo{author}{Vinyals, O.} \&
  \bibinfo{author}{Dean, J.}
\newblock \bibinfo{title}{Distilling the knowledge in a neural network}.
\newblock \emph{\bibinfo{journal}{arXiv:1503.02531}}  (\bibinfo{year}{2015}).

\bibitem{Chiaradia_2020}
\bibinfo{author}{Chiaradia, I.} \& \bibinfo{author}{Lancaster, M.~A.}
\newblock \bibinfo{title}{Brain organoids for the study of human neurobiology
  at the interface of in vitro and in vivo}.
\newblock \emph{\bibinfo{journal}{Nat. Neurosci.}}
  \textbf{\bibinfo{volume}{23}}, \bibinfo{pages}{1496--1508}
  (\bibinfo{year}{2020}).

\bibitem{Kim_2020}
\bibinfo{author}{Kim, J.}, \bibinfo{author}{Koo, B.-K.} \&
  \bibinfo{author}{Knoblich, J.~A.}
\newblock \bibinfo{title}{Human organoids: model systems for human biology and
  medicine}.
\newblock \emph{\bibinfo{journal}{Nat. Rev. Mol. Cell Biol.}}
  \textbf{\bibinfo{volume}{21}}, \bibinfo{pages}{571--584}
  (\bibinfo{year}{2020}).

\bibitem{Liu_2021}
\bibinfo{author}{Liu, M.}, \bibinfo{author}{Fang, S.}, \bibinfo{author}{Dong,
  H.} \& \bibinfo{author}{Xu, C.}
\newblock \bibinfo{title}{Review of digital twin about concepts, technologies,
  and industrial applications}.
\newblock \emph{\bibinfo{journal}{J. Manuf. Syst.}}
  \textbf{\bibinfo{volume}{58}}, \bibinfo{pages}{346--361}
  (\bibinfo{year}{2021}).

\bibitem{wiki_Self_organization}
\bibinfo{title}{Self-organization.}
  \urlprefix\url{https://en.wikipedia.org/wiki/Self-organization}.

\bibitem{wiki_Spontaneous_order}
\bibinfo{title}{Spontaneous order.}
  \urlprefix\url{https://en.wikipedia.org/wiki/Spontaneous_order}.

\bibitem{Cambria_2014}
\bibinfo{author}{Cambria, E.} \& \bibinfo{author}{White, B.}
\newblock \bibinfo{title}{Jumping {NLP} curves: A review of natural language
  processing research}.
\newblock \emph{\bibinfo{journal}{IEEE Comput. Intell. Mag.}}
  \textbf{\bibinfo{volume}{9}}, \bibinfo{pages}{48--57} (\bibinfo{year}{2014}).

\bibitem{Goldberg_2017}
\bibinfo{author}{Goldberg, Y.}
\newblock \emph{\bibinfo{title}{Neural Network Methods in Natural Language
  Processing}} (\bibinfo{publisher}{Morgan \& Claypool Publishers},
  \bibinfo{year}{2017}).

\bibitem{Goodfellow_2016}
\bibinfo{author}{I.~Goodfellow, Y.~B.} \& \bibinfo{author}{Courville, A.}
\newblock \emph{\bibinfo{title}{Deep learning}} (\bibinfo{publisher}{The MIT
  Press}, \bibinfo{address}{Cambridge}, \bibinfo{year}{2016}).

\bibitem{Stanley_2019}
\bibinfo{author}{Stanley, K.~O.}, \bibinfo{author}{Clune, J.},
  \bibinfo{author}{Lehman, J.} \& \bibinfo{author}{Miikkulainen, R.}
\newblock \bibinfo{title}{Designing neural networks through neuroevolution}.
\newblock \emph{\bibinfo{journal}{Nat. Mach. Intell.}}
  \textbf{\bibinfo{volume}{1}}, \bibinfo{pages}{24--35} (\bibinfo{year}{2019}).

\bibitem{Stanley_2015}
\bibinfo{author}{Stanley, K.~O.} \& \bibinfo{author}{Lehman, J.}
\newblock \emph{\bibinfo{title}{Why Greatness Cannot Be Planned: The Myth of
  the Objective}} (\bibinfo{publisher}{Springer}, \bibinfo{year}{2015}).

\bibitem{wiki_selective_breeding}
\bibinfo{title}{Selective breeding.}
  \urlprefix\url{https://en.wikipedia.org/wiki/Selective_breeding}.

\bibitem{Davis_2017}
\bibinfo{author}{Davis, A.~M.}, \bibinfo{author}{Plowright, A.~T.} \&
  \bibinfo{author}{Valeur, E.}
\newblock \bibinfo{title}{Directing evolution: the next revolution in drug
  discovery?}
\newblock \emph{\bibinfo{journal}{Nat. Rev. Drug Discov.}}
  \textbf{\bibinfo{volume}{16}}, \bibinfo{pages}{681--698}
  (\bibinfo{year}{2017}).

\bibitem{Romero_2009}
\bibinfo{author}{Romero, P.} \& \bibinfo{author}{Arnold, F.}
\newblock \bibinfo{title}{Exploring protein fitness landscapes by directed
  evolution}.
\newblock \emph{\bibinfo{journal}{Nat. Rev. Mol. Cell Biol.}}
  \textbf{\bibinfo{volume}{16}}, \bibinfo{pages}{866--876}
  (\bibinfo{year}{2009}).

\bibitem{Subramanian_2020}
\bibinfo{author}{Subramanian, I.}, \bibinfo{author}{Verma, S.},
  \bibinfo{author}{Kumar, S.}, \bibinfo{author}{Jere, A.} \&
  \bibinfo{author}{Anamika, K.}
\newblock \bibinfo{title}{Multi-omics data integration, interpretation, and its
  application}.
\newblock \emph{\bibinfo{journal}{Bioinform Biol. Insights}}
  \textbf{\bibinfo{volume}{14}}, \bibinfo{pages}{1177932219899051}
  (\bibinfo{year}{2020}).

\end{thebibliography}

\end{document}